\documentclass[sigconf]{acmart}
\usepackage{graphicx}
\usepackage{multirow}
\usepackage{amsmath}
\usepackage{colortbl}
\usepackage{fontawesome5}

\AtBeginDocument{%
  }

\copyrightyear{2026}
\acmYear{2026}
\setcopyright{cc}
\setcctype{by-nc-nd}
\acmConference[CHI '26]{Proceedings of the 2026 CHI Conference on Human Factors in Computing Systems}{April 13--17, 2026}{Barcelona, Spain}
\acmBooktitle{Proceedings of the 2026 CHI Conference on Human Factors in Computing Systems (CHI '26), April 13--17, 2026, Barcelona, Spain}
\acmPrice{}
\acmDOI{10.1145/3772318.3790965}
\acmISBN{979-8-4007-2278-3/2026/04}

\begin{document}

\title[Sensing Your Vocals]{Sensing Your Vocals: Exploring the Activity of Vocal Cord Muscles for Pitch Assessment Using Electromyography and Ultrasonography}

\author{Kanyu Chen}
\orcid{0000-0001-5095-533X}
\affiliation{%
  \institution{Keio University}
  \city{Yokohama}
  \country{Japan}
}
\email{cady.cky@kmd.keio.ac.jp}

\author{Rebecca Panskus}
\orcid{0009-0006-9751-675X}
\affiliation{%
  \institution{Ruhr University Bochum}
  \city{Bochum}
  \country{Germany}
}
\email{rebecca.panskus@rub.de}

\author{Erwin Wu}
\orcid{0000-0002-6723-2864}
\affiliation{%
  \institution{Institute of Science Tokyo}
  \city{Tokyo}
  \country{Japan}
}
\email{wu.e.aa@vogue.cs.titech.ac.jp}

\author{Yichen Peng}
\orcid{0000-0002-8544-3905}
\affiliation{%
  \institution{Institute of Science Tokyo}
  \city{Tokyo}
  \country{Japan}
}
\email{peng.y.ag@m.titech.ac.jp}

\author{Daichi Saito}
\orcid{0000-0003-3843-7542}
\affiliation{%
  \institution{Institute of Science Tokyo}
  \city{Tokyo}
  \country{Japan}
}
\email{saito.d.ah@m.titech.ac.jp}

\author{Emiko Kamiyama}
\orcid{0009-0009-2772-2437}
\affiliation{%
  \institution{Keio University}
  \city{Yokohama}
  \country{Japan}
}
\email{emiko.pooh@kmd.keio.ac.jp}

\author{Ruiteng Li}
\orcid{0009-0002-4970-2145}
\affiliation{%
  \institution{Waseda University}
  \city{Tokyo}
  \country{Japan}
}
\email{ruiteng.li@toki.waseda.jp}

\author{Chen-Chieh Liao}
\orcid{0000-0002-9850-2468}
\affiliation{%
  \institution{Institute of Science Tokyo}
  \city{Tokyo}
  \country{Japan}
}
\email{liao.c.aa@m.titech.ac.jp}

\author{Karola Marky}
\orcid{0000-0001-7129-9642}
\affiliation{%
  \institution{Ruhr University Bochum}
  \city{Bochum}
  \country{Germany}
}
\email{karola.marky@rub.de}

\author{Kato Akira}
\orcid{0000-0002-6129-7747}
\affiliation{%
  \institution{Keio University}
  \city{Yokohama}
  \country{Japan}
}
\email{kato@kmd.keio.ac.jp}

\author{Hideki Koike}
\orcid{0000-0002-8989-6434}
\affiliation{%
  \institution{Institute of Science Tokyo}
  \city{Tokyo}
  \country{Japan}
}
\email{koike@c.titech.ac.jp}

\author{Kai Kunze}
\orcid{0000-0003-2294-3774}
\affiliation{%
  \institution{Keio University}
  \city{Yokohama}
  \country{Japan}
}
\email{kai@kmd.keio.ac.jp}

\renewcommand{\shortauthors}{Kanyu CHEN et al.}

\begin{abstract}
Vocal training is difficult because the muscles that control pitch, resonance, and phonation are internal and invisible to learners. This paper investigates how Electromyography (EMG) and ultrasonic imaging (UI) can make these muscles observable for training purposes. We report three studies.
First, we analyze the EMG and UI data from 16 singers (beginners, experienced \& professionals), revealing differences among three vocal groups of the muscle control proficiency. Second, we use the collected data to create a system that visualizes an expert's muscle activity as reference. This system is tested in a user study with 12 novices, showing that EMG highlighted muscle activation nuances, while UI provided insights into vocal cord length and dynamics. Third, to compare our approach to traditional methods (audio analysis and coach instructions), we conducted a focus group study with 15 experienced singers. Our results suggest that EMG is promising for improving vocal skill development and enhancing feedback systems. We conclude the paper with a detailed comparison of the analyzed modalities (EMG, UI and traditional methods), resulting in recommendations to improve vocal muscle training systems.
\end{abstract}



\maketitle

\section{INTRODUCTION}
\label{sec:1 Introduction}
\begin{figure*}
  \centering
  \includegraphics[width=\textwidth]{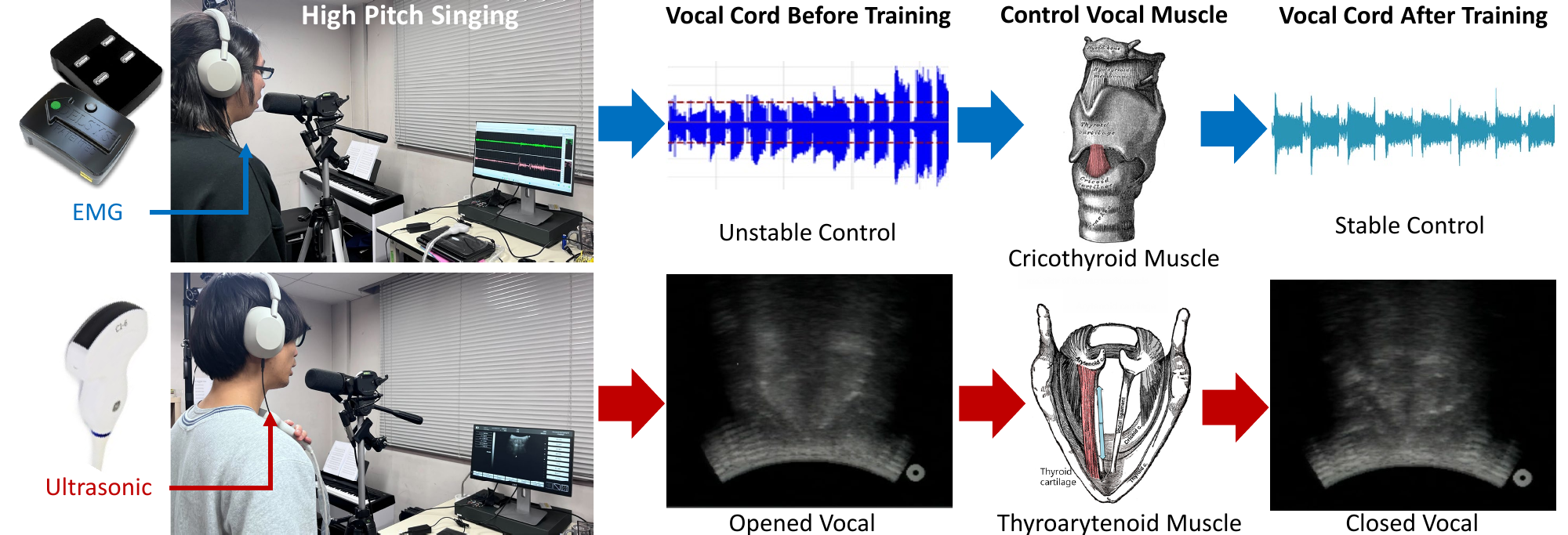}
  \caption{From sensor data collection to visualization: (1) Showing a participant wearing an EMG sensors to detect the activity of the cricothyroid muscle. (2) Showing a participant using a B-type ultrasonography device to visualize the thyroarytenoid muscle.}
  \Description{This figure gives an overview of how singing data was collected within our study once with EMG and once with ultrasonography sensors.
  Furthermore, graphical visualizations of this data are presented. The examples show the difference of the stability control and vocal cords in opened and closed position pre and post training.}
  \label{fig:teaser}
\end{figure*}

Vocal training depends on precise control of muscles that most people cannot see or feel directly. Singers must learn to coordinate the thyroarytenoid and cricothyroid muscles to control pitch, yet they typically receive feedback only through sound-based methods such as spectrograms~\cite{luck2018comparison} or, in clinical settings, through invasive laryngoscopy~\cite{song2013assessment}. Previous studies have demonstrated the value of metrics like the Singing Power Ratio (SPR) in distinguishing vocal capabilities~\cite{usha2017objective} and have explored respiratory muscle training to enhance vocal function~\cite{desjardins2022respiratory}. These approaches remain limited, however, because they provide no intuitive connection between what singers hear and what their muscles are doing. Spectrograms show acoustic output but not the physical mechanisms producing it. Singers are left to discover through trial and error how to engage their vocal muscles to achieve a target sound.

This paper addresses these gaps by investigating electromyography (EMG) and ultrasonographic imaging (UI) as basis for more intuitive and less invasive visualization methods for vocal cords training. EMG and UI are methods to measure muscle movements~\cite{buchthal1959electromyography, kimura2019sottovoce}. Tension of muscles~\cite{titze2002rules}, such as the thyroarytenoid and cricothyroid muscles, allows pitch control in voice production. Our goal is leveraging the understanding of muscle activity to improve the visualization of vocal activity. For this, we conducted a series of studies with singers since singing is a specific skill that requires the control of vocal cord muscles.

We conducted three studies to evaluate EMG and UI for vocal training. First, we collected a Vocal Cord Sensing Dataset (VCSD) from 16 singers with varying skill levels, using both EMG and UI to establish baseline patterns of vocal muscle activity. This dataset addresses our first research question:

\textbf{RQ1:} Can EMG and UI be used to capture different levels of proficiency in singing and support singers of various skill levels?

Analysis of the dataset revealed distinct muscle activity patterns between experienced and novice singers, confirming that both modalities can detect proficiency differences. Having established a technical foundation, we next used the data of experienced singers to create reference visualizations for EMG and UI of ``correct'' singing that novice singers could use for training. In our second study, we tested whether novice singers could use these references to improve their own performance:

\textbf{RQ2:} How do EMG and UI perform in training novice singers?

We recruited 12 novice singers to practice matching their muscle activity to the EMG and UI reference visualizations in real time. Results showed that both modalities support skill development, though with different strengths. EMG feedback enhanced perceived controllability and reduced fatigue, while UI feedback improved vocal cord control for specific pitches but required higher cognitive effort. The UI setup also constrained movement due to its physical size, limiting its use to stationary practice.

Having EMG and UI tested for novices, we created a more mobile setup based on EMG-only and wireless microphones to support performance-oriented scenarios where singer might move. UI was not considered in this study due to its rather bulky setup and the already established muscle control of experienced singers. Hence, in our final study with 15 experienced singers, we wanted to shed light on the following research question:

\textbf{RQ3:} How does a portable EMG setup perform in supporting experienced singers compared to traditional feedback methods?

Our findings showed that EMG captured nuanced muscle engagement, with tenor singers aligning closer to expert references. In addition, we found a strong correlation with vocal power, or known as Singing Power Ratio (SPR), highlighting superior muscle control in experts. Comparing our sensing setup with traditional methods, we found that EMG can provide technical precision and contextual guidance, showcasing its potential as a portable tool for advanced vocal training. Yet, overall a combination of EMG and traditional methods is likely to yield the best results as both approaches complement each other.

Across all three studies, we found that EMG and UI feedback can be effective for vocal training. While our investigation focused on singing, the underlying techniques have broader applications in speech detection, language learning, and voice capture. We conclude with guidelines for designing effective technology-assisted practice systems.\vspace{2mm}

\noindent\textbf{Contribution.}
The major contributions of this paper are:\begin{enumerate}
    \item \textbf{Improving vocal pitch assessment:} We propose an accessible approach for collecting and evaluating vocal pitch using EMG and UI sensing based on real-world data.
    \item \textbf{In-depth investigation with singers:} We present three user studies (in total N=43) varying untrained (novice) and trained groups to further investigate the effectiveness, usability, skill transferability of vocal sensing techniques and provide a comprehensive design guideline for vocal training systems. We provide an open-access dataset of dual-modality sensing (EMG and ultrasonography) collected from 16 participants with varying levels of vocal training experience, available under an osf link: \url{https://osf.io/gkjsx/?view_only=0e6aee05b63040a5b1f140297775ece9}.
    \item \textbf{Broadening EMG and UI for other domains:} We show how our results motivate and inform further research in related domains of vocal training also providing actionable guidelines for effective technology-assisted practice systems.
\end{enumerate}

\section{RELATED WORK}
\label{sec: 2 Related Work}
This paper builds on four aspects below, using Electromyography (EMG) and ultrasonic imaging (UI) to explore how physiological data can improve vocal skill assessment.

\subsection{Vocal Sensing and Feedback in Interactive Music Systems}
\label{sec: 2.1 Interactive Systems for Music Perception and Learning}
Interactive music systems have shown potential for musical perception, bodily awareness, and learning outcomes. For example, the integration of digital tools and games~\cite{10.1145/2468356.2468435} to support the voice treatment of parkinson patients demonstrates how gamification can improve vocal therapy by increasing patient participation and vocal loudness. Similarly, Bean Academy~\cite{10.1145/3544549.3583824} simplifies music composition through vocal query transcription, reducing the barriers to music theory comprehension and software proficiency.

Research on vocal sensing technologies, such as pitch detection, remains in early stages. Celestia~\cite{10.1145/2468356.2479485} uses pitch detection for real-time interaction in music games, showing the potential of auditory feedback to improve vocal control. Additionally, BrainiBeats~\cite{10.1145/3544549.3585910} uses physiological signals (EMG and EEG) to generate music based on emotional states, merging creativity with biological data. These systems suggest applications for personalized learning environments and therapeutic interventions.

Research has integrated physiological responses into artistic performances~\cite{han2022linking,he2022frisson}, and interactive installations have further explored embodied music creation. ``The Music Room''~\cite{10.1145/2468356.2479620} is an immersive installation where participants' movements within the physical space trigger and shape musical elements, demonstrating collaborative, embodied music-making. Similarly, ``The Throat III''~\cite{10.1145/2468356.2479596} and ``The Vocal Corder''~\cite{10.1145/2556288.2557050, 10.1145/2559206.2574798} allow opera singers to manipulate their voices dynamically, expanding the expressive possibilities of vocal performance through sensor-based technology. Mixed reality technologies~\cite{10.1145/3544548.3581162} have also shown a significant increase in co-presence for remote collaboration with musical partners, offering opportunities for vocal sensing.

\subsection{Physiological Foundations of Vocal Pitch Control}
\label{sec: 2.2 Integrating Vocal Theory with Technological Advancements}
Vocal training combines foundational techniques with physiological understanding to optimize performance. Estill Voice Training (EVT)~\cite{Steinhauer2019}, Speech Level Singing (SLS)~\cite{mcclellan2011comparative}, and Complete Vocal Technique (CVT)~\cite{Sundberg2017} represent three widely-used methods. Traditional approaches to vocal assessment focus on breath control, pitch accuracy, volume modulation, and rhythmic precision~\cite{welch2019oxford}.

In this paper, our sensing methods to pitch assessment is built on EVT theory~\cite{Steinhauer2019}, which suggests that vocal pitch is determined by the tension in the vocal folds, a factor evident in both speech and song~\cite{belyk2018does, titze2002rules}. Previous studies identified the thyroarytenoid and cricothyroid muscles as the primary controllers of vocal pitch~\cite{titze2002rules}, with their contractions adjusting the tension of the vocal folds by rocking the cricoid and arytenoid cartilages. This mechanism alters the pitch of the voice~\cite{buchthal1959electromyography, gay1972electromyography,kempster1988effects}.

However, a persistent challenge in vocal pedagogy~\cite{naseth2012constructing} is the lack of comprehensive feedback mechanisms, particularly in the absence of continuous guidance from professional instructors. Muscle tension is central to discussions of vocal performance pedagogy. Controlled tension is necessary for precise manipulation of the vocal tract, but excessive tension -- is often associated with reduced sound quality~\cite{miller1996structure}. Current methods~\cite{naseth2012constructing} predominantly rely on auditory feedback and visual analysis of spectrograms to evaluate vocal performance, with limited insights into the underlying muscular mechanisms involved in sound production. Emerging trends in vocal training research~\cite{chen2024novel, chen2025multimodal, chen2025exploring, welch2019oxford} have begun to integrate technological innovations, such as muscle activity visualization and monitoring, to offer more precise and informative feedback to learners.

Therefore, to integrate the sensing technologies in vocal assessment can enhance the efficiency and efficacy of vocal training programs, considering the diverse needs and skill levels of practitioners across various musical genres and educational settings.

\subsection{Sensing Technologies for Vocal Analysis}
\label{sec: 2.3 Emerging Input Techniques in Vocal Sensing}
Recent advancements in sensing technologies improve the understanding and control of the human voice. Traditional vocal analysis methods~\cite{naseth2012constructing} often rely on subjective assessments, which can be difficult to obtain consistently and are prone to interpretation biases. For instance, waveform analysis can detect timing errors and improve note alignment, articulation, and transitions, enhancing recording quality~\cite{allen2019advanced}. Recent improvements also enable more detailed processing of complex vocal performances~\cite{huang2021enhanced}. Singing Power Ratio (SPR) has been used to provide automated feedback in vocal training, assessing resonance and clarity, crucial in classical singing~\cite{lee2022real, sundberg2020singing}. Despite these advancements, traditional methods still limit a deeper understanding of vocal production mechanisms.

EMG and ultrasonography have enabled researchers to examine vocal physiology in greater detail. Existing research has explored correlations between uttering vowels or sentences and engagement of speech muscles~\cite{zhu2022towards}, the use of pitch and EMG for omohyoid detection~\cite{vojtech2021surface}, and analysis of face and neck muscle movements during speech~\cite{reed2020surface}. Ultrasonic techniques capture speech dynamics from tongue movement to sound production~\cite{kimura2019sottovoce}.

Certain wearable designs worn on the throat or chest offer a potential avenue for enhancing bodily signals as input for vocal expression, from a custom wearable collar that monitors vocal performance~\cite{reed2022singing}, to a breath-related wearable instrument that enhances somatic awareness (the conscious perception of internal bodily sensations) during singing~\cite{Cotton2021Body}. Despite significant progress, a fully integrated sensing system specifically tailored for detecting vocal pitch muscles remains absent. Nevertheless, the integration of EMG and ultrasound into practical, wearable systems for real-time vocal analysis is still an area ripe for exploration, with promising applications in both clinical settings and vocal pedagogy.

\subsection{Muscle Activity Sensing: From General Applications to Vocal Training}
\label{sec: 2.4 Input Sensing in Muscle Activity Detection and Visualization Beyond Vocal Muscles}
The assessment of muscle engagement has increasingly captivated the Human-Computer Interaction (HCI) community~\cite{Eddy2023}, particularly within the realms of motor skill training and rehabilitation~\cite{Saisho2019, Papakostas2019}, owing to its potential to elevate skill acquisition and optimize performance outcomes.
Ultrasound stands out as a traditional yet potent modality for visualizing muscle movements.
As a medical imaging technique, it furnishes images of internal bodily structures, including muscles~\cite{Sarto2021-zb, Nagae2023-ql}, thereby offering users a tangible means to visually comprehend muscle movements.
In this study, we leverage ultrasonography imaging (UI) to provide real-time visualization of vocal cord movements as a training method.

However, the interpretation of UI can present challenges, and quantifying these images' outcomes often entails complexity~\cite{Paris2021-ck,meng2025placebo, meng2023towards,han2022linking}.
In contrast, Electromyography (EMG), Mechanomyogram (MMG), and Electrical Impedance Tomography (EIT) present alternative methodologies for measuring muscle activity.
EMG, by recording the action potentials generated by muscle contractions as a result of neural activity~\cite{lopes2015,takahashi20221}, offers a high degree of sensitivity to the electrical activities underlying muscle contractions.
This attribute renders EMG particularly adept at capturing the subtle nuances in vocal cord muscle movements, which are characterized by rapid and complex motion.
MMG, in contrast, detects mechanical signals from muscle surfaces during contraction, through technologies such as capacitive plane arrays~\cite{rudolph2022} and force-sensitive resistors~\cite{Esposito2018-mu}.
However, given its propensity for monitoring mechanical vibrations, MMG is more apt for assessing larger muscle groups rather than the fine movements of the vocal cords.

Electrical Impedance Tomography (EIT) has been explored for capturing hand gesture movements~\cite{Zhang2016, Zheng2021} and muscle activity~\cite{Zhu2021,Zhu2022}. Unlike EMG and MMG, EIT monitors both contracted and stretched muscles by mapping internal conductivity changes, providing regional and cross-sectional information. However, its precision may not meet the requirements for vocal cord training, where targeted monitoring of specific muscles is essential. EMG sensors provide localized measurement, making them effective for isolating the activity of small or deeply located muscles such as those controlling the vocal cords.

For vocal training applications requiring precise feedback on small, rapidly moving muscles, EMG offers advantages as a measurement technique. Its ability to provide immediate, specific data on muscle activity makes it valuable for detailed analysis and feedback on vocal performance. Combined with ultrasound visualization, these complementary modalities can address gaps in current vocal training feedback systems.

\section{Research Approach \& Hardware Setup}
\label{methodology overview}
Our research program comprised three sequential user studies, each building upon findings from its predecessor. We first describe the iterative investigation structure, then detail the hardware configurations employed across studies.

\subsection{Iterative Investigations}

We conducted three studies to systematically examine EMG and ultrasound imaging (UI) as modalities for vocal training feedback. Each study addressed distinct research questions while informing subsequent methodological refinements.

\begin{itemize}
    \item \textbf{Study 1 -- Data Collection (N=16)}: This foundational study investigated whether EMG and UI can effectively visualize vocal cord muscle activity and differentiate between skill levels~\ref{sec:4 VOCAL CORD SENSING DATASET}. We recruited 16 singers spanning novice to professional expertise to establish reference datasets. The study addressed two questions: (a) Do measurable differences in vocal cord activity patterns exist between differently skilled singers? (b) Can these differences be rendered as meaningful visualizations? Findings from this study informed sensor selection and feedback design for subsequent investigations.

    \item \textbf{Study 2 -- Novice Study (N=12)}: In the second study, we specifically focused on novice singers to investigate their temporal stability through EMG and feedback on muscle movements by UI also capturing the task load and usability of the setup. In this study, we used a static setup since the training of basic skills does not require much movement.

    \item \textbf{Study 3 -- Experienced Study(N=15)}: In our final study, we investigated a more realistic setting for advanced singers requiring a mobile setup that captures the muscle activity of singing during other movements. Because of that, we had to rule out UI sensors, since they are rather bulky and require the singers to be in a static position. Further, we used this study to gather feedback from the experienced singers that compares our setup with traditional coaching.
\end{itemize}

\begin{figure*}
    \centering
    \includegraphics[width=\textwidth]{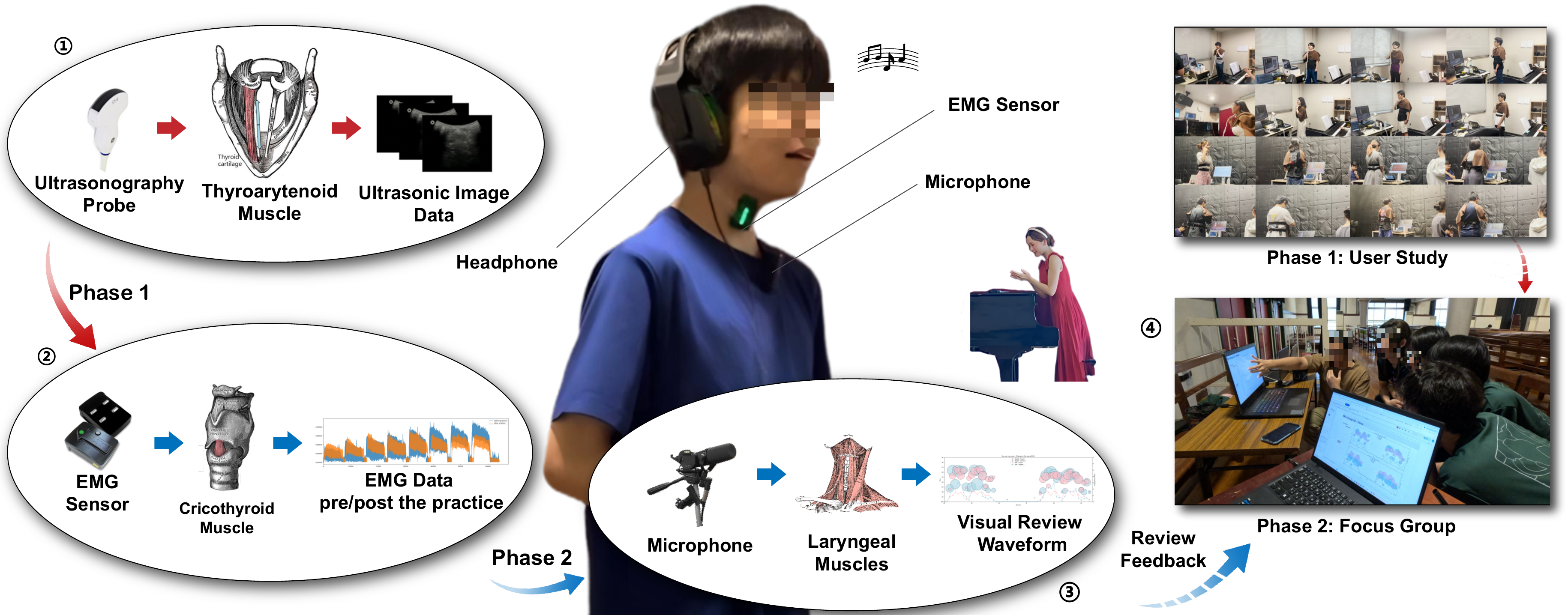}
    \caption{Overview of the Iterative Study Process and the Hardware Setup: (1) Ultrasonography sensors require the manual positioning of the probe on the skin near the vocal cords. (2) EMG sensors need to be placed bilaterally around the user's larynx to monitor vocal muscle activity. (3) The data recording also includes the use of a noise-canceling microphone.}
    \label{fig:overveiw}
    \Description{The figure illustrates the iterative study process and the hardware setup, comprising EMG and Ultrasonography sensing for the vocal cord activity: (1) For ultrasonography sensors the manual positioning of the probe on the skin near the vocal cords is required. (2) EMG sensors need to be placed bilaterally around the user's larynx to monitor vocal muscle activity. (3) The data recording also includes the use of a noise-canceling microphone.}
\end{figure*}

\subsection{Hardware Setup}
\label{hardware setup}
Based on previous work, we considered the thyroarytenoid and cricothyroid muscles as the primary controllers of vocal pitch~\cite{titze2002rules}. We collected two types of data to directly measure muscle activity across various pitches: EMG and UI. A microphone collected audio.

\noindent\paragraph{\textbf{EMG Setup}}
To capture EMG, we used Delsys Trigno Wireless EMG sensors to measure vocal muscle activity. Sensors were positioned on the anterior neck between the thyroid cartilage (Adam's apple) and the superior border of the cricoid cartilage. The data was captured at 2000Hz in the first study (VCSD dataset) and second study (novice study). In the third study (experienced singers study), EMG data was sampled at 4370Hz to segment short-duration pitch events during song passages, and to align with the higher-resolution audio.

\noindent\paragraph{\textbf{Ultrasonography}}
The CONTEC CMS600P2 Full Digital B-type Ultrasound Diagnostic System was employed. Participants held the probe at the front of the larynx, with ultrasonic operating at a 3.5 MHz transmit frequency and the system recording video at 30 fps. Ultrasonography was used only in the novice study for visual feedback on vocal muscle activity. Due to its bulkiness, this method was not used in the experienced singers study.

\noindent\paragraph{\textbf{Audio Capture}}
Study 2 employed a Shure SM7B microphone for high-fidelity recording in controlled acoustic conditions. Study 3 substituted a portable wireless microphone to accommodate singer movement during performance tasks.

\subsection{Data Analysis Statistical Rationale}
All statistical analyses were conducted in Python using \texttt{numpy}, \texttt{scipy}, and \texttt{statsmodels}. Parametric (\textit{t}-tests) or non-parametric (Wilcoxon signed-rank test) tests were chosen accordingly based on normality
assumptions. Multiple comparisons were corrected using the Benjamini--Hochberg FDR procedure~\cite{benjamini1995controlling}. Effect sizes (Cohen's $d$ or rank-biserial $r_{\mathrm{rb}}$) and 95\% confidence intervals were reported following recommended standards~\cite{cohen2013statistical,
lakens2013calculating}. For repeated-measures outcomes (e.g., pitch-wise EMG stability and vocal-fold length), linear mixed-effects models~\cite{gelman2007data} were used with participant as a random intercept and Group and Pitch as fixed effects.

\subsection{Ethical Considerations}\label{sec:ethics}
Our institutional Ethical Review Board (ERB) approved our study design. Throughout our study, we took precautions to treat our participants in an ethically correct manner and adhered to strict privacy laws. Participants of all three studies received a consent form containing detailed information about the captured data, assurances that they could withdraw from participating at any point without negative consequences, and information about their rights in compliance with the General Data Protection Regulation (GDPR) and national data protection laws. Participating in any of the three studies caused no risks beyond everyday life or psychological harm for participants. To protect the privacy of our participants and their environment, we have anonymized the transcripts prior to analysis as well as the photos of our participants used in this paper.

\subsection{Positionality Statement}
This project was a collaboration between HCI and security and privacy (S\&P) researchers from academia and industry. All members of the research team have more than three years of experience in conducting, writing, publishing, and reviewing HCI research, while focusing on end users, children, and singers. We approach this study as researchers located within the HCI and S\&P communities. Our interest in this topic stems from our own experience conducting qualitative HCI and S\&P research with end users, children, and singers, where we have identified, first-hand, several constraints and challenges and aim to understand the extent to which these constraints and challenges occur on a scale. Nine of the authors grew up in non-Western countries but now collaborate with Western institutions. Three authors grew up in Western countries. The resulting research experiences with end users and children from different cultural backgrounds shape our understanding of these target groups. We acknowledge that our dual roles as conducting this study and as fellow researchers in this space may have influenced how participants shared their experiences, while also enabling us to engage with them more meaningfully. At the same time, our positions as HCI and S\&P researchers limit our ability to interpret findings beyond the perspectives of our own discipline. Our vision for this work is to serve as a starting point for our research communities to work together in addressing the challenges of conducting inclusive research and enacting changes.

\section{STUDY 1: INITIAL EMG \& UI INVESTIGATION (RQ1)}
\label{sec:4 VOCAL CORD SENSING DATASET}
The primary goal of this study was to conduct an exploratory analysis of vocal muscle activity during singing using EMG and UI, investigating whether these sensing technologies can capture different levels of singing proficiency. To this end, we collected a Vocal Cord Sensing Dataset (VCSD).

\subsection{Collected Data}
\label{sec: 4.1 Dataset Contents}
The collected VCSD dataset consists of a total recording of around three hours. The details of the dataset can be found in Table~\ref{tab:data}. Besides the videos of ultrasonography and the 2-channel EMG data, the dataset also includes the breath sensor values and the reference videos with audio during the collection procedure.

\begin{table*}[]
\resizebox{\textwidth}{!}{%
\begin{tabular}{@{}ccccccccccccccccccc@{}}
\toprule
\textbf{Dataset}                                                            & \textbf{Sampling}                                      & \multicolumn{10}{c}{\textbf{Novices (1-10)}}                                                                                                                                                                                                                                                                                                                                                                                                                                                                                                                                         & \multicolumn{3}{c}{\textbf{\begin{tabular}[c]{@{}c@{}}Experienced \\  (11-13)\end{tabular}}}                                                                   & \multicolumn{3}{c}{\textbf{Professionals (14-16)}}                                                                                                                              & \textbf{Total Size} \\ \cmidrule(lr){1-1} \cmidrule(lr){2-2} \cmidrule(lr){3-12} \cmidrule(lr){13-15} \cmidrule(lr){16-18} \cmidrule(lr){19-19}
Subject                                                                     & -                                                      & 1                                                      & 2                                                     & 3                                                     & 4                                                      & 5                                                      & 6                                                      & 7                                                      & 8                                                      & 9                                                      & 10                                                     & 11                                                     & 12                                                     & 13                                                     & 14                                                     & 15                                                     & 16                                                     & 16                  \\
\begin{tabular}[c]{@{}c@{}}Pitch Range (the\\ number of pitch)\end{tabular} & \begin{tabular}[c]{@{}c@{}}G2 - E6\\ (27)\end{tabular} & \begin{tabular}[c]{@{}c@{}}F3 - F4\\ (14)\end{tabular} & \begin{tabular}[c]{@{}c@{}}C3 - C4\\ (8)\end{tabular} & \begin{tabular}[c]{@{}c@{}}F3 - G4\\ (9)\end{tabular} & \begin{tabular}[c]{@{}c@{}}C3 - B4\\ (14)\end{tabular} & \begin{tabular}[c]{@{}c@{}}G3 - D5\\ (12)\end{tabular} & \begin{tabular}[c]{@{}c@{}}G3 - D5\\ (12)\end{tabular} & \begin{tabular}[c]{@{}c@{}}D3 - C5\\ (14)\end{tabular} & \begin{tabular}[c]{@{}c@{}}G3 - C5\\ (11)\end{tabular} & \begin{tabular}[c]{@{}c@{}}F3 - D5\\ (13)\end{tabular} & \begin{tabular}[c]{@{}c@{}}F3 - D5\\ (13)\end{tabular} & \begin{tabular}[c]{@{}c@{}}F3 - E5\\ (14)\end{tabular} & \begin{tabular}[c]{@{}c@{}}F2 - C5\\ (19)\end{tabular} & \begin{tabular}[c]{@{}c@{}}G2 - E5\\ (20)\end{tabular} & \begin{tabular}[c]{@{}c@{}}E3 = E6\\ (22)\end{tabular} & \begin{tabular}[c]{@{}c@{}}D3 - E6\\ (22)\end{tabular} & \begin{tabular}[c]{@{}c@{}}G2 - C6\\ (25)\end{tabular} & G2 - E6             \\
\textbf{\begin{tabular}[c]{@{}c@{}}2-channel\\ EMG Data\end{tabular}}       & 2000 hz/s                                              & 316s                                                   & 253s                                                  & 179s                                                  & 383s                                                   & 348s                                                   & 277s                                                   & 342s                                                   & 130s                                                   & 245s                                                   & 329s                                                   & 298s                                                   & 249s                                                   & 421s                                                   & 139s                                                   & 495s                                                   & 524s                                                   & 4928s               \\
\textbf{\begin{tabular}[c]{@{}c@{}}Ultrasonography\\ Data\end{tabular}}     & 30 fps                                                 & 273s                                                   & 333s                                                  & 264s                                                  & 287s                                                   & 217s                                                   & 213s                                                   & 280s                                                   & 336s                                                   & 232s                                                   & 288s                                                   & 333s                                                   & 288s                                                   & 368s                                                   & 483s                                                   & 583s                                                   & 360s                                                   & 5138s               \\ \bottomrule
\end{tabular}%
}
\vspace{3mm}
\caption{Statistics of the VCSD datasets: EMG, UI, and pitch range for 16 users across skill levels (Novice, Intermediate, Expert)}
\Description{This table details the dataset description, including 2-channel EMG (2000 hz), ultrasound data (30 fps) and pitch range for 16 users. (Novices (10), Experienced (3) and Professionals (3)). EMG data acquisition ranged from 130s -524s, total time 4928s. Ultrasonography data acquisition ranged from 213s -583s, total time 5138s. The pitch range length of the tones collected by the subjects from 1 to 16 are [14, 8, 9, 14, 12, 12, 14, 11, 13, 13, 14, 11 , 13, 13, 14, 19, 20, 22, 22, 25], with a total range of G2 to E6 and 27 piano white key pitches.}
\label{tab:data}
\end{table*}

\begin{figure*}
    \centering
    \includegraphics[width=\textwidth]{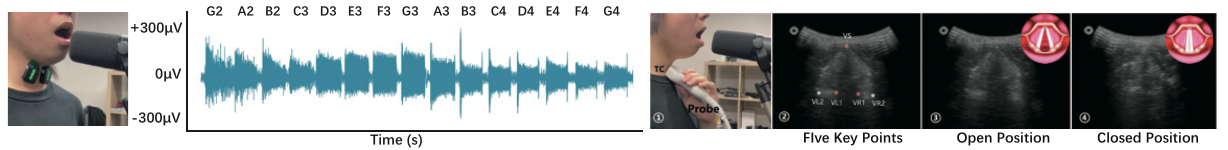}
    \caption{Left: EMG sensor placement and visualization of a sample of raw EMG data accompanied by muscle/cartilage position annotations.  Right: Positioning of the ultrasonography probe and sample of raw ultrasound imaging data accompanied by muscle/cartilage position annotations.}
    \Description{Samples of raw EMG data and ultrasound imaging data are presented in this figure, each accompanied by annotations indicating muscle and cartilage positions. In the EMG data, muscle and cartilage positions are highlighted, while in the ultrasound imaging, corresponding annotations are provided.}
    \label{fig:raw_raw}
\end{figure*}

\subsection{Study Procedure}
\label{sec 4.3 Collection Procedure}

\noindent\textit{1) Welcome \& Familiarization}: Before data collection, the participants were informed about their rights and the implications of their participation. When agreeing to participation and use of their data, they signed a consent form. Then, an initial questionnaire was given to collect the participants' demographics and their vocal training experience.
Next, participants were introduced to the sensing devices and the task, followed by a 5-minute practice session to familiarize themselves with the sensing devices and the task.\newline

\noindent\textit{2) Singing Task}: Participants were instructed to sing their vocal range as accurately as possible, referenced the scientific pitch notation (SPN). To support pitch accuracy, an 80 bpm piano reference of the scale was played in real time. Participants were asked to follow this reference closely and maintain each pitch for two seconds. Each participant performed the task with both sensing modalities (EMG and UI). Each session was repeated four times, yielding eight recording rounds per participant. Session order was counterbalanced across participants to mitigate learning effects.  \newline

\noindent\textit{3) Exit Interview}: Finally, the participants were interviewed about vocal training and their impressions of the two sensing technologies.

\subsection{Participants \& Recruitment}
\label{sec: 4.2 Participants}
We recruited 16 participants (6 female, 10 male, aged 21-33 years, mean=25.7)
from two local institutes. Ten participants were beginners with little vocal training experience, and three participants were intermediate amateurs who had basic vocal knowledge. The remaining three participants were experts who experienced professional vocal training for more than 10 years.
Novice and experienced participants were provided a 1000 yen gift card per hour as reimbursement for their participation in the study. Expert participants were reimbursed with 10,000 yen as compensation for their time and effort.
The collection process was approved by the IRB of the local institutes. All data collected within the scope of this study are anonymous and are only released with the approval of the participants.

\subsection{Results}
\label{sec: 4.4 Analysis and Insights}
Since the raw data contains noise and redundant information, we first post-processed the data and analyzed the results from different perspectives, to obtain insights from the collected data.
\subsubsection{Data Processing}
\paragraph{{EMG}}\label{sec:4.1.1}
Since our focus was on the stability and controllability of the cricothyroid muscle, we tried to extract stability information from the data. The raw data was first denoised through a moving average filter (window size = 10ms), then a Hilbert transform was performed to calculate envelopes of the signal~\cite{cohen1995time, oppenheim1999discrete}. The stability of muscle activity~\cite{farrus2007jitter} $s$ was then calculated as follows:

\begin{equation}
\begin{split}
    s = \frac{1}{N-1}\sum_{t=1}^{N-1}\|20log\frac{A_{t+1}}{A_t}\|
\end{split}
\label{eq:1}
\end{equation}

$A_t$ denotes the previous envelope value of the filtered EMG at timestep $t$.
This equation is designed with reference to shimmer measurement in voice, which is frequently used in the field of acoustic analysis~\cite{teixeira2015acoustic}.
Dividing among envelopes allows calculating the stability of each pitch and comparing the differences in scales between each participant.

\paragraph{{UI}}\label{sec:4.1.2}
The UI video had to be quantified for analysis. Therefore, we developed a landmark detector for tracking the vocal cord muscle from the ultrasonic images. Here, we focused on five important key points in the video: start points (connection) of two vocal cords, ends of the inner side of vocal cords, and the end of the outer side of vocal cords, as shown in Figure~\ref{fig:raw_raw}. These five points discern changes in the true vocal cord structure and cartilage position based on previous research~\cite{kumar2017vocal}.
Since the shape of the vocal cord differs for each participant, we manually annotated the key points on an initial frame for each session.
With the positional data from the five key points, we computed the length of the true vocal cords (depicted in red in Figure~\ref{fig:raw_raw}) as follows:

\begin{equation}
\begin{split}
    L = \frac{1}{2} *  \left( Dist \left( P_{VS} , \dfrac{P_{VL1} + P_{VL2}}{2}\right) +  Dist \left( P_{VS} , \dfrac{P_{VR1} + P_{VR2}}{2}\right)\right)
\end{split}
\end{equation}

\subsubsection{Data Analysis}
\label{sec: 4.4.2 Data Analysis}
\begin{figure*}
    \centering
    \includegraphics[width=\textwidth]{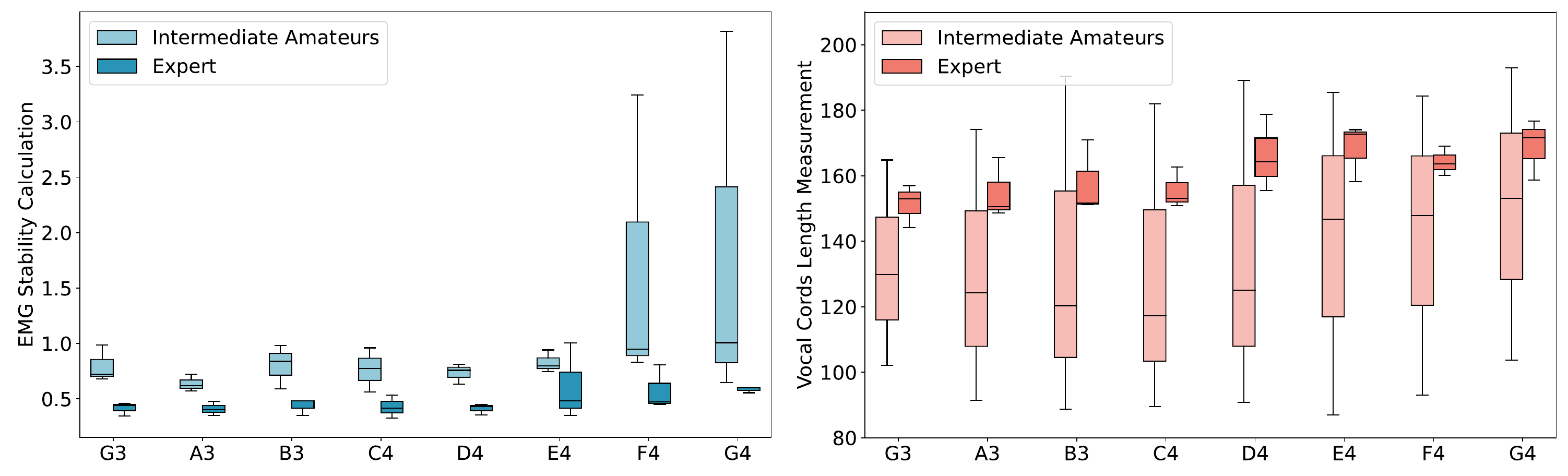}
    \caption{Comparison of the stability calculation derived from EMG data and the measured vocal cord length via ultrasonography across a pitch range from G3 to G4. Left a set of a participant's raw EMG imaging data, right a set of a participant's ultrasound imaging data, each spanning over one octave.}
    \Description{This figure presents a boxplot comparing stability calculations from EMG data and vocal cord lengths measured via ultrasonography across the pitch range from G3 to G4. The data highlight distinctions between intermediate and expert skill levels, with experts showing a narrower range of values compared to intermediates. Additionally, the boxplot reveals a trend of increasing average vocal cord length with higher pitches, particularly pronounced among expert participants.}
    \label{fig:fig56}
\end{figure*}
To understand the effect of the two sensing data as well as to answer RQ1, we further analyzed the collected data among the participants.

Following recommendations for statistical reporting
~\cite{cohen2013statistical, lakens2013calculating}, we report effect sizes
(partial $\eta^2$) alongside $p$-values. First, the processed data was statistically analyzed using linear mixed-effects models (LME)~\cite{gelman2007data} among the three groups of different level participants (beginners, intermediate amateurs, experts). \textit{Pitch} was treated as a repeated factor, \textit{Group} of different levels as a between-subject factor, and \textit{Gender} as a covariate. Participant number was included as a random intercept to account for within-subject dependency across pitches. Since most beginners cannot ``correctly'' sing the notes, we focused on the data between the intermediate and expert-level participants in the following analysis.

\paragraph{EMG}
\label{sec:4.1.1}
For the stability score, we picked up the common range (G3-G4) of the intermediate and expert groups to perform a deeper investigation. The mixed-effects analysis revealed a significant main effect of Group, with experts showing markedly lower EMG instability than intermediate singers ($\beta = -0.48$, 95\% CI~$[-0.83, -0.12]$, $p = .013$, partial $\eta^2 = .18$). Because higher EMG stability scores indicate poorer temporal muscle control, this result reflects substantially more stable cricothyroid activity among expert singers across the octave. The main effect of Pitch was not significant ($p = .28$), although EMG instability tended to increase at higher notes (F4--G4). The Group~$\times$~Pitch interaction was not significant ($p = .62$), suggesting a consistent stability advantage for experts across the octave range. Gender had no significant influence on EMG ($p = .26$).

The results per pitch are shown in the left figure of Figure~\ref{fig:fig56}. Despite an overall better temporal stability of the expert group, we found an obvious difference in the higher pitches (F4 and G4), which indicates the ability to control vocal muscles in high-pitch sounds of the experts.

\paragraph{UI}
\label{sec:4.1.2}
For the vocal length data estimated from the UI, there was no significant difference between intermediate and expert singers ($\beta = -1.82$, 95\% CI~$[-6.97, 3.34]$, $p = .49$). Pitch did not reach statistical significance ($p = .29$), although an expected increase in length was observed toward higher notes. Gender showed a significant main effect ($\beta = 12.43$, 95\% CI~$[4.85, 20.01]$, $p = .003$, partial $\eta^2 = .25$), indicating systematic differences in UI measurements between male and female singers. No significant Group~$\times$~Pitch interaction was observed ($p = .99$).

The results suggest a similar trend as the EMG data that the range of vocal cords is more stable for the experts (see right part of Figure~\ref{fig:fig56}). Although the UI does not provide any temporal information, it shows that expert singers can more precisely manage their vocal cords.

\paragraph{Interpretation Across Pitches}
As visualized in Figure~\ref{fig:fig56}, expert singers demonstrate lower EMG instability (better temporal muscle control), particularly at higher pitches. They also show more consistent vocal cord positioning in the UI data, indicated by narrower distributions of measured lengths. Intermediate singers, in contrast, exhibit greater variability in both EMG and UI, highlighting less stable laryngeal control strategies.

\subsubsection{Result Summary: RQ1}
In sum, EMG and UI can both effectively capture the vocal cord muscle activity and be used to differentiate the level of singers in the scale task.
The EMG feedback was easy to understand and provided a better representation of the temporal stability of vocal cords. UI offers a direct visual cue of vocal cords, providing intuitive feedback of muscle activities.
\vspace{5mm}

\section{STUDY 2: NOVICE STUDY (RQ2)}
\label{sec:5 NOVICE USER STUDY}
\begin{figure*}[htbp]
    \centering
    \includegraphics[width=\textwidth]{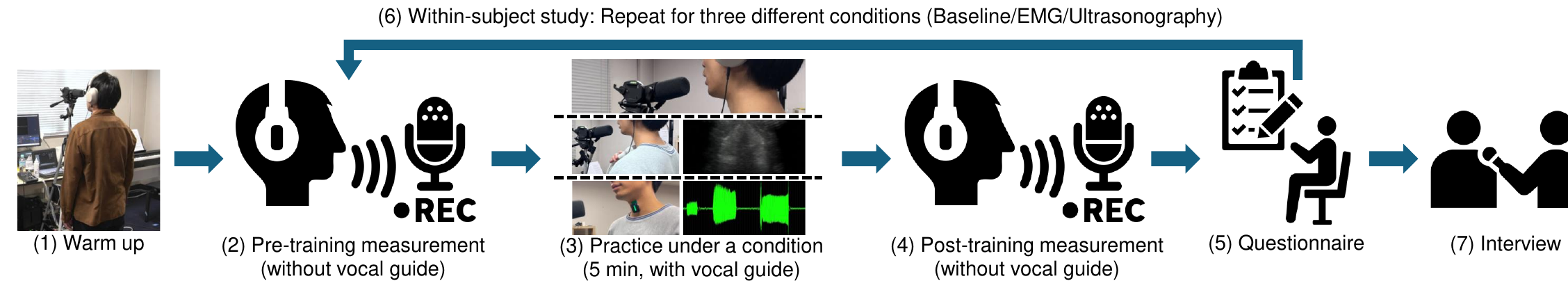}
    \caption{The procedure of the novice study.}
    \Description{This figure shows that the novice study followed a structured procedure: warm-up, pre-training measurements (including EMG and ultrasonography), guided practice, post-training assessments, questionnaire completion, and interviews for feedback.}
    \label{fig:procedure}
\end{figure*}

Study 1 demonstrated that EMG and UI can differentiate vocal cord muscle activity across skill levels, with EMG capturing temporal stability and UI providing direct visual feedback of vocal cord positioning. Building on these findings, we conducted a study with novice singers to evaluate how each feedback modality supports vocal training. Participants practiced matching their muscle activity to expert reference visualizations derived from the Vocal Cord Sensing Dataset (VCSD) collected in Study 1.

\subsection{Study Procedure}

The study employed a within-subjects design in which each participant completed training under three conditions: audio-only baseline, EMG feedback, and UI feedback. Condition order was counterbalanced across participants. The procedure (see Figure~\ref{fig:procedure}) consisted of five phases:

\noindent\textit{1) Welcome \& Familiarization}: First, we welcomed the participants and obtained their consent.\newline

\noindent\textit{2) Warm-Up \& Pre-Training}: We did a warm-up for 5-10 minutes to prepare the participants for the vocal exercise. Next, we measured a pre-training baseline for EMG and UI.\newline

\noindent\textit{3) Practice Sessions}: The participants practiced for 5 minutes under one of the three conditions with a vocal guide which is brief demonstration on vocal muscle structure and the vocalis mechanism. The participants then learned from experts' reference recording (that we obtained from the VCSD collected in Study 1) for each practice session.\newline

\noindent\textit{4) Post-Training}: Next, we measured the post-training performance without a vocal guide and asked the participants to fill in a questionnaire about the condition. To capture the perceived task load, we used the NASA-TLX~\cite{hart1988development}. We further captured a subset of the Sense of Agency Scale (SoAS)~\cite{soas2017}. Here, we extracted five pertinent questions from the SoAS, focusing on evaluating aspects of controllability. The steps (2) to (5) were repeated for three conditions. The order of EMG and UI was counterbalanced.\newline

\noindent\textit{5) Exit Interview}: The participants were interviewed about their preferences, evaluation, and suggestions for the system. Further, we captured the participants' feedback on the advantageous and less beneficial aspects of each training methodology.

\subsection{Participants \& Recruitment}
To investigate the training effects of visual cues from both above vocal cords sensing, 12 novice participants (6 male, 6 female; aged 24-34 years, mean=27.4) were recruited.
Regarding the singing (vocal training) experience, none of the participants had experienced professional training before, while all participants are singing enthusiasts who visit Karaoke (or similar activities) at least once a month.
Participants received a 1000 yen gift card per hour as reimbursement for their participation in the study. The study was approved by the local Institutional Review Board.

\begin{figure*}
    \centering
    \includegraphics[width=\linewidth]{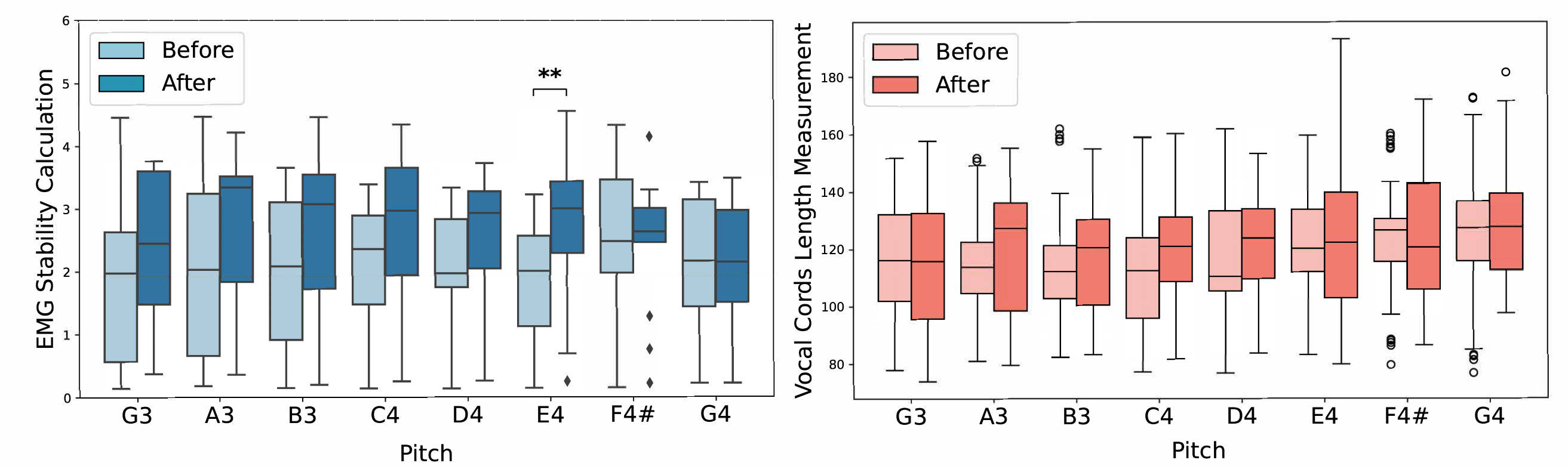}
    \caption{Left: Results of the stability $s\downarrow$ test on the vocal muscle activity following the Equation~\ref{eq:1}, pre- ("before" in light blue) and post-training ("after" in dark blue). Right: Results of the manually annotating on the vocal length form ultrasonography according to instruction in Section \ref{sec:4.1.2}, pre- ("before" in light red) and post-training ("after" in dark red). In both graphs the Wilcoxon's test results are indicated by the bridges between boxes (**: $p < 0.01$).}
    \Description{This figure showcases boxplots representing the stability of EMG signals across various pitches before and after practice, alongside measurements of vocal cord length obtained from ultrasonography. It reveals a decrease in the overall stability of EMG signals post-practice, indicated by higher stability value. Additionally, the range of vocal cord length increases following practice.}
    \label{fig:emg_us}
\end{figure*}

\vspace{3mm}
\subsection{Results}
\label{sec: 5.2 Results}

\subsubsection{\textbf{Sensing:}
\label{sec: 5.2.1 Sensing}
EMG Stability Decreased, While Mean Vocal Cords Lengths Increased After the Practice.}

\begin{figure*}
    \centering
    \includegraphics[width=0.8\textwidth]{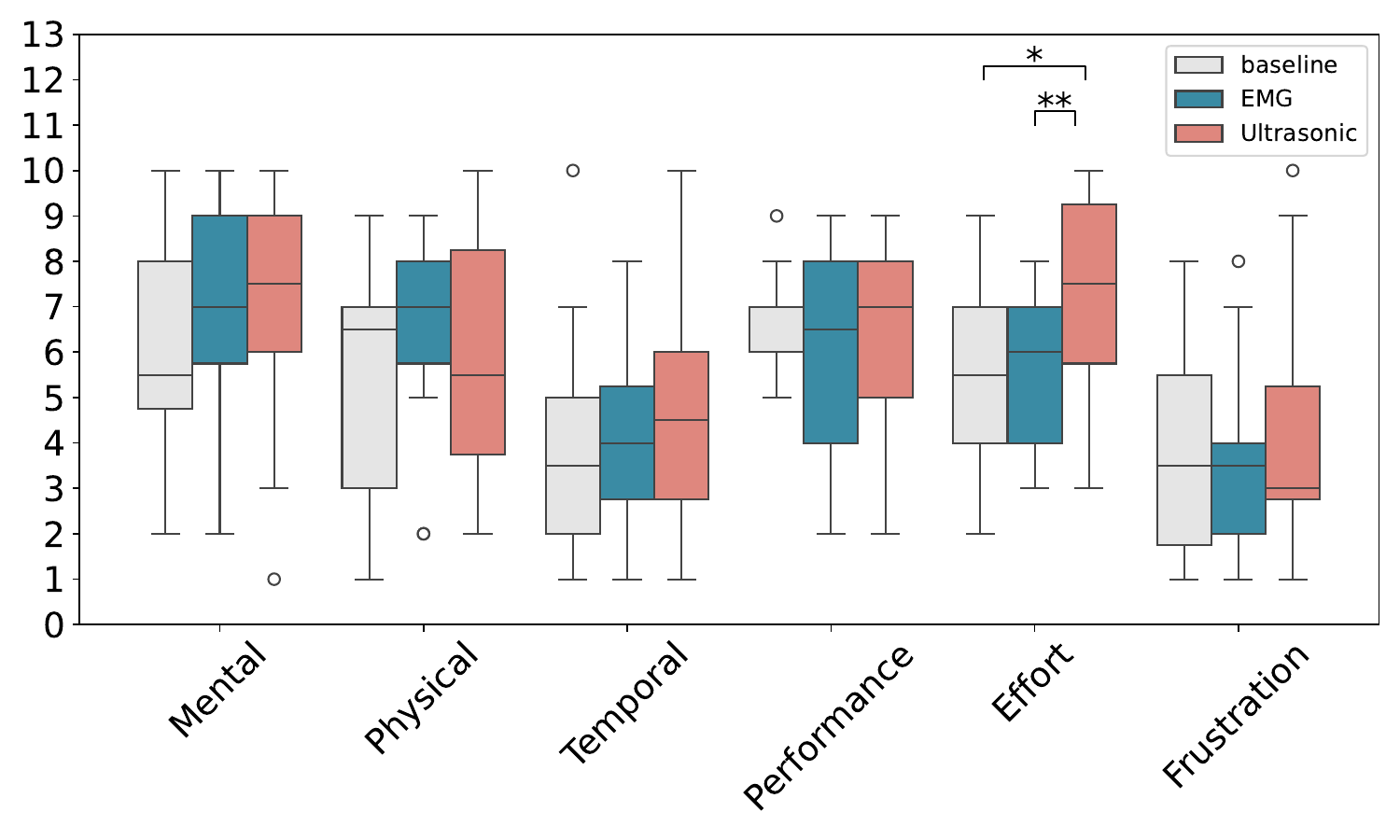}
    \caption{NASA-TLX Scores Comparison Across Baseline, EMG, and UI Methods with the Wilcoxon signed-rank test (*: $p < 0.05$, **: $p < 0.01$).}
    \Description{This figure presents the scores obtained from the NASA-TLX questionnaire across three methods: baseline, EMG, and ultrasonography. It delineates a marginal increase in scores for ultrasonography compared to EMG, with a further slight elevation above baseline levels. Notably, the dimension of "effort" exhibits a statistically significant disparity between ultrasonography and EMG, with a p-value of 0.0441. Moreover, the contrast between ultrasonography and baseline is more pronounced, yielding a p-value of 0.0094.}
    \label{fig:nasatlx}
\end{figure*}

\paragraph{Stability Controlling (EMG)}
We evaluated muscle activity during vocal training, calculating the stability $s\downarrow$ of EMG signals (see Section~\ref{sec:4.1.1}) before and after practicing with the EMG training method. For each pitch (G3–G4), we conducted paired Wilcoxon
signed-rank tests to compare pre–post stability within participants and applied Benjamini-Hochberg false discouvery rate (FDR) correction~\cite{benjamini1995controlling}. Before correction, three pitches (B3, C4, E4) showed significant increases in
instability after training (B3: $W = 6$, $p = .048$,
$r_{\mathrm{rb}} = .62$, 95\% CI $[.08, .88]$; C4 ($W = 8$, $p = .065$,
$r_{\mathrm{rb}} = .47$, 95\% CI $[-.15, .82]$);
E4: $W = 14$, $p = .009$, $r_{\mathrm{rb}} = .74$, 95\% CI $[.22, .93]$). After applying FDR correction, only E4 remained statistically significant
($p_{\mathrm{FDR}} = .036$), while the remaining pitches did not survive
correction ($p_{\mathrm{FDR}} > .10$). The post-training stability was generally lower than the pre-training levels (see Figure~\ref{fig:emg_us} (left)). This finding stands in contrast to the self-reported perceptions of increased controllability as indicated in the responses to the adapted SoAS questionnaire. Notably, this decrease in stability was statistically significant, especially for pitches \textit{B3}, \textit{C3}, and \textit{E4}, with $W = 6, p = 0.048$, $W = 3,p = 0.049$, and $W =14, p = 0.009$ respectively, as determined by the Wilcoxon's signed-rank test.

To address potential confounding factors, we fitted a linear mixed-effects model~\cite{gelman2007data} with fixed factors Pre/Post, Pitch, Order, and Gender, and random intercepts for participants. The model revealed a significant main effect of Pre/Post, indicating increased instability after training ($\beta = 0.41$, 95\% CI $[0.12, 0.69]$, $p = .018$, partial $R^2 = .21$). Order ($\beta = 0.08$, 95\% CI $[-0.11, 0.27]$, $p = .41$) and Gender
($\beta = 0.10$, 95\% CI $[-0.12, 0.32]$, $p = .33$) were not significant. No significant interactions were observed.

The decrease in EMG stability could be attributed to distraction, as participants primarily focused on maintaining stable visual feedback while managing the cognitive demands of interpreting raw EMG data. We further address this in the discussion section.

\paragraph{Vocal Lengths Controlling (UI)}
We analyzed UI recordings by annotating five frames per pitch (compared to one frame in Study 1) to increase measurement reliability. Mean vocal cord length increased for each pitch after UI-feedback training (Figure~\ref{fig:emg_us}, right panel).
Before multiple-comparison correction, we observed medium-to-large effects and trends toward significance at pitches B3 ($W = 3$, $p = .089$, $r = .49$, 95\% CI $[-.12, .83]$) and C4 ($W = 7$, $p = .058$, $r = .52$, 95\% CI $[-.08, .84]$), and a significant increase at D4 ($W = 4$, $p = .049$, $r = .57$, 95\% CI $[-.01, .86]$). However, after controlling the false discovery rate across all eight pitches using the Benjamini--Hochberg procedure~\cite{benjamini1995controlling}, none of these effects remained statistically significant (B3: $p_{\mathrm{FDR}} = .24$, C4 and D4: $p_{\mathrm{FDR}} = .23$). Despite the lack of FDR-corrected significance, the consistent positive effect sizes suggest a systematic tendency toward longer vocal cord configurations after training, which aligns with the pattern observed in the VCSD, where expert singers exhibited longer and more stable vocal cord lengths than intermediate participants.

\subsubsection{\textbf{Perception:} EMG Enhanced Perceived Controllability, While UI Increased Mental Workload.}
\label{sec: 5.2.2 Perception}

\paragraph{NASA-TLX}
We compared the workload between three methods: (1) training with audio-only feedback (baseline), (2) training with audio and EMG visual feedback, and (3) training with UI feedback (see Figure~\ref{fig:nasatlx}). Our results show that both the training with EMG and UI feedback have a slightly higher workload than the baseline. Further, training with UI has a slightly higher workload than training with EMG.

When comparing UI against the Baseline condition, we observed medium-to-large effects for Performance ($W = 6$, $p = .071$, $r_{\mathrm{rb}} = .56$, 95\% CI $[-.09, .86]$, $p_{\mathrm{FDR}} = .14$) and Effort ($W = 14$, $p = .044$, $r_{\mathrm{rb}} = .63$, 95\% CI $[.03, .88]$, $p_{\mathrm{FDR}} = .10$), with the latter approaching statistical significance
after correction. Other dimensions (Mental, Physical, Temporal, Frustration) showed smaller effects and did not reach significance ($p_{\mathrm{FDR}} > .30$).
Comparing UI to EMG revealed a similar pattern: Effort showed a large and statistically significant increase for UI ($W = 1.5$, $p = .0094$, $r_{\mathrm{rb}} = .71$, 95\% CI $[.18, .92]$, $p_{\mathrm{FDR}} = .028$),
while Mental and Physical workloads showed moderate but non-significant trends ($p_{\mathrm{FDR}} = .18$ and $.13$). No significant differences were observed for Temporal, Performance, or Frustration ($p_{\mathrm{FDR}} > .30$).
It is expected that both of our sensing methods offer participants a direction of active practice to use their vocal muscles to sing, instead of adjusting the way to pronounce passively based on the pitch from audio feedback. Participants perceived a greater improvement in their vocal \textit{Performance} following training with the proposed systems, as compared to solely relying on auditory feedback from hearing their own audio. This subjective assessment underscores the enhanced efficacy of the training methodologies implemented in our study.

\paragraph{Controllability}
\begin{figure*}
    \centering
    \includegraphics[width=0.8\textwidth]{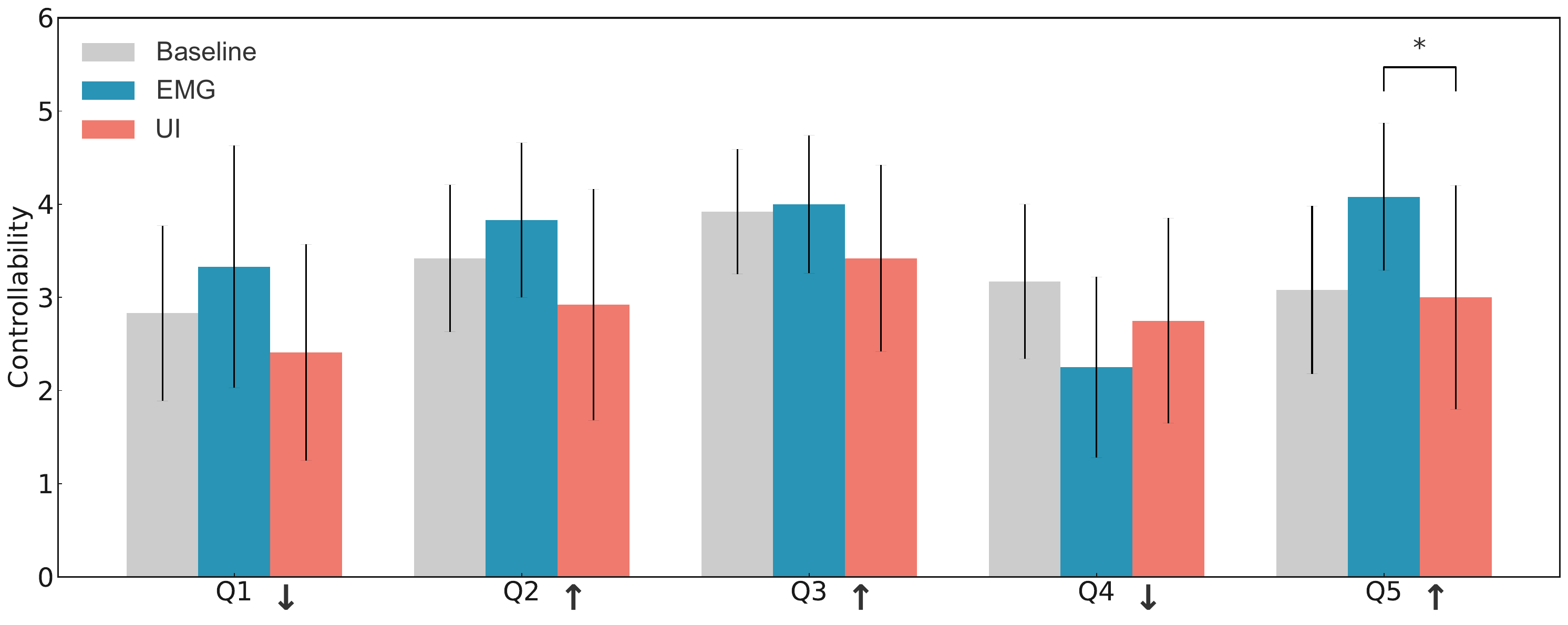}
    \caption{The results of the pick-up SoAS questionnaire with the Wilcoxon signed-rank test. (*: $p < 0.05$. Error bars represent standard deviations. Above questions (Q1-Q5) lists as following: (Q1) My actions just happen without my intention. (Q2) My behavior is planned by me from the very beginning to the very end. (Q3) I am completely responsible for everything that results from my actions. (Q4) My movements are automatic - my body simply makes them. (Q5) I am in full control of what I do.}
    \Description{This figure shows the results of the pick-up SoAS questionnaire alongside the Wilcoxon signed-rank test results. It highlights participant perceptions from the adapted SoAS questionnaire, and significant differences were found between EMG and Ultrasonography methods, notably in Q1 (p = 0.0482), Q2 (p = 0.0451), and Q5 (p = 0.0125).}
    \label{fig:controllability}
\end{figure*}

The analysis of the adapted SoAS questionnaire (see Figure~\ref{fig:controllability}), indicates an increased perception of controllability among participants using our proposed vocal muscle sensing system, as compared to three alternative training methodologies. Before correction~\cite{benjamini1995controlling}, we compared EMG-based feedback
against UI feedback using Wilcoxon signed-rank tests that participants showed significantly higher controllability ratings for EMG in Q1 ($W = 3$, $p = .048$, $r_{\mathrm{rb}} = .67$, 95\% CI $[.06, .91]$), Q2 ($W = 3$, $p = .045$, $r_{\mathrm{rb}} = .68$, 95\% CI $[.07, .91]$), and Q5 ($W = 7$, $p = .013$, $r_{\mathrm{rb}} = .74$, 95\% CI $[.22, .93]$). After FDR correction, the effect for Q5 remained significant ($p_{\mathrm{FDR}} = .032$), while Q1 and Q2 showed marginal trends ($p_{\mathrm{FDR}} = .072$). Q3 and Q4 did not show significant differences ($p_{\mathrm{FDR}} > .40$), consistent with their smaller effect sizes ($r_{\mathrm{rb}} < .30$). Notably, the EMG approach demonstrated superior controllability, possibly because of the more intuitive representation of EMG signals. Furthermore, the EMG methodology notably outperformed in \textit{Q5}, which pertains to self-presentation confidence~\cite{soas2017}.

\begin{figure*}
    \centering
    \includegraphics[width=\textwidth]{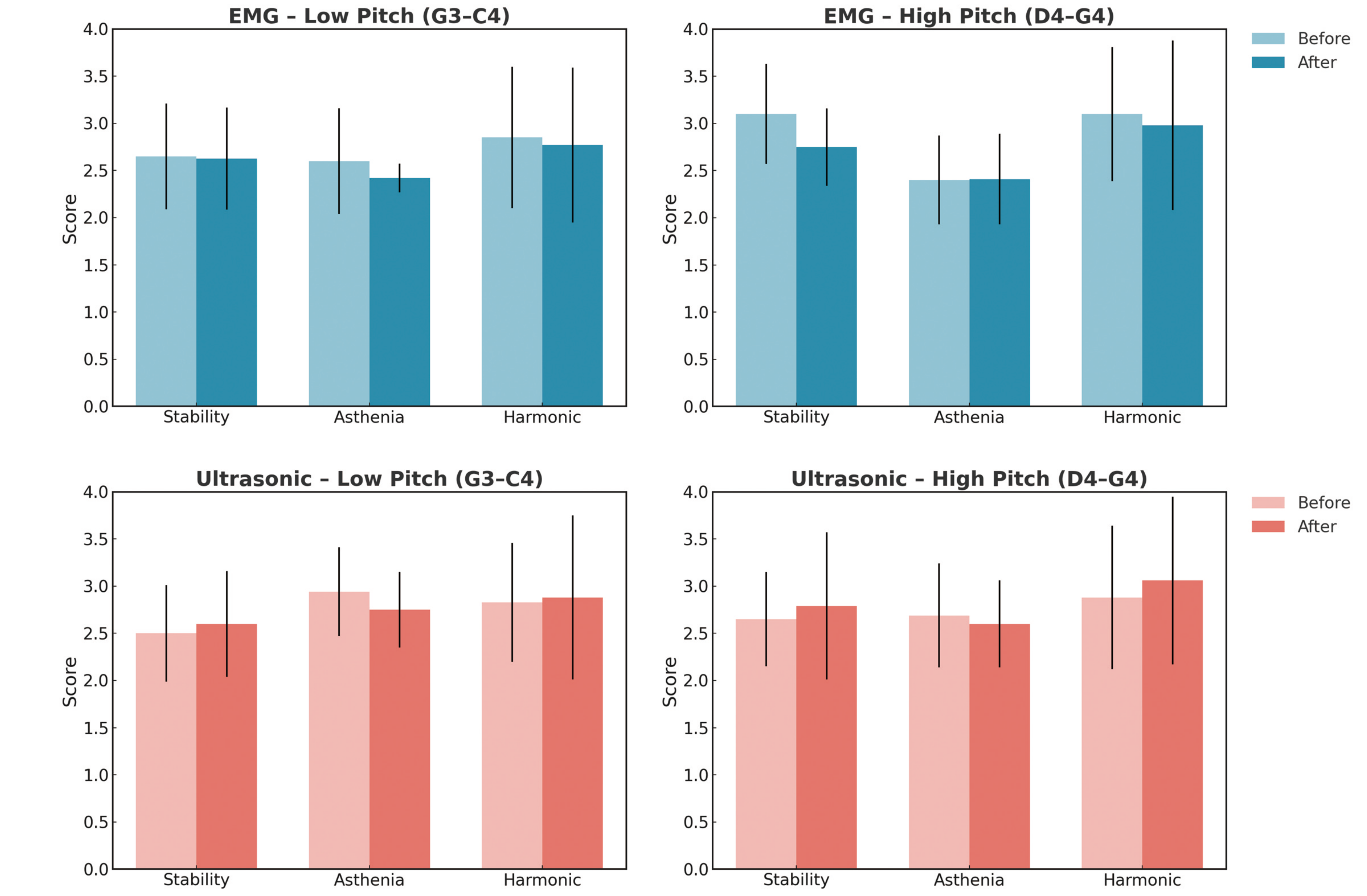}
    \caption{The outcomes of the expert evaluations regarding participants' singing performances are quantified in three distinct aspects: Stability$\uparrow$, Asthenia$\downarrow$, and Harmonic$\uparrow$, shown in the bar chart above.}
    \Description{This bar chart detailed the expert evaluations of participants' low and high pitch singing performances are quantified across three distinct aspects: Stability, Asthenia, and Harmonic quality. Ultrasonic assessments are highlighted with green boxes, indicating an improvement in each performance post-practice. Conversely, EMG evaluations are represented with red boxes, signifying a decline in the respective aspect, except for asthenia, which improved in low pitch performance.}
    \label{fig:ratings}
\end{figure*}

\subsubsection{\textbf{Audience Involvement:} EMG Enhances Control and Reduces Fatigue; UI Visualizes Muscles but Lacks Precision}
\label{sec: 5.2.3 Audience Invovlement}
Audience involvement integrates expert evaluations and performer self-assess- \\ ments to provide a comprehensive evaluation of vocal performance and the usability of sensors. Therefore, we first invited an expert group for scoring participants recorded audio across our three practice methods. Second, participants engaged in a playback session of their own recordings during an exit interview, incorporating their reflections into the evaluation.

\paragraph{\textbf{Expert Group}}
Here is a brief description about three experts' background and their scoring criteria.
\begin{itemize}
    \item [Exp1:] 5 years experience in music production and piano practice since childhood, this expert focused on evaluating both bass and tenor recordings, noting fluctuations in performance due to vocal fatigue.
    \item [Exp2:] An acoustic engineer with 15 years of piano practice and vocal knowledge, this expert prioritized vocal stability, accuracy, and loudness, emphasizing the impact of gender and vocal range on performance.
    \item [Exp3:] A former professional musical stage actor with over 10 years of experience, this expert assessed recordings based on vocal stability and pitch control, noting difficulties in low notes, particularly for female participants.
\end{itemize}

\paragraph{Rating of Experts} To ensure reliability, all audio evaluations were conducted on both pre- and post-training recordings using consistent recording equipment and environmental settings.

In our investigation, the expert evaluations focused on scoring participants recorded audio to appraise participants vocal performances separating in low\&high pitch conditions of participants before\&after they underwent training based on three critical aspects: Stability, Asthenia, and Harmonic. According to the data provided in Figure~\ref{fig:ratings}, the experts observed that vocal performances enhanced across all three aspects following training with the UI method. In contrast, training with the EMG method yielded somewhat diminished performance levels, as per expert assessments. Notably, the scores reflecting improvement in high-pitched vocalizations were more pronounced compared to those in lower pitches, exhibiting average enhancements of approximately 0.157 and 0.113 respectively, on a 5-point likert scale (see Appendix B). This scale was developed through expert discussions and references to previous research~\cite{yamauchi2010perceptual, de1997test}.

\paragraph{Participants and Experts Interviews}
\textit{Practice Feedback Suggestion from Experts:} Experts observed notable differences in the performance of vocal stability, harmonics, and asthenia across the training methods. Exp2 rated audio (condition 1 with expert's reference practice) and UI (condition 3 with ultrasonograph practice) feedback higher, stating, ``\textit{1, 3 sound clearer and louder than 2 [with EMG practice]; there are fewer noises, less hoarseness and bubbling sounds.}'' (Exp2). Exp3, however, noted the consistent difficulties participants had with low notes: ``\textit{Females generally have difficulty in stably pronouncing the low notes... it may be better to change the key for female practice.}'' (Exp3).\\

\textit{Participants' Reported Feedback on Sensing Methods.} From the participant feedback, EMG proved more effective in enhancing control and reducing fatigue. As P6 noted, ``\textit{Following the chart helps improve vocal position in real-time, and reduces vocal fatigue.}'' (P6). Participants also highlighted how EMG offered real-time insights, with P7 commenting on the teacher's vocal stability: ``\textit{I noticed that the teacher's vocal EMG changes were stable during vocalization.}'' (P7).

Conversely, UI provided valuable visualizations but was less intuitive for control. P4 mentioned, ``\textit{Ultrasound clearly shows the muscle movement process.}'' but P8 found it challenging to control: ``\textit{I find it hard to control my vocals to the exact position I want.}''. Exp2 confirmed this, stating, ``\textit{Ultrasonography sounds it takes slightly more efforts than EMG.}'' (Exp2).

Participants found the baseline method to rely heavily on understanding the expert's reference.
As P3 said, ``\textit{I try to adjust pitch based on reference sound feedback.}'', indicating reliance on auditory feedback rather than active vocal control. EMG, on the other hand, facilitated real-time correction, while UI offered clear visualizations but lacked precision for improving vocal positioning.

\subsubsection{Result Summary: RQ2}
In sum, EMG feedback significantly enhanced the participants' perceived sense of control and self-presentation confidence, outperforming both the baseline and UI feedback. Despite this increased perception, actual muscle stability decreased post-practice, possibly due to cognitive overload from visual feedback interpretation. In contrast, UI feedback led to measurable improvements in vocal control, particularly in lengthening vocal cords for specific pitches, although some participants struggled with precise control due to usability challenges with the UI probe. Overall, both methods showed benefits, but with different strengths and challenges in practice.

\section{STUDY 3: EXPERIENCED SINGERS STUDY (RQ3)}
\label{sec:6 Trained Group Study}
In our final study, we investigated the perceptions of experienced singers to investigate RQ3: How does a portable EMG setup perform in supporting experienced singers compared to traditional feedback methods?

\begin{figure*}
    \centering
    \includegraphics[width=\textwidth]{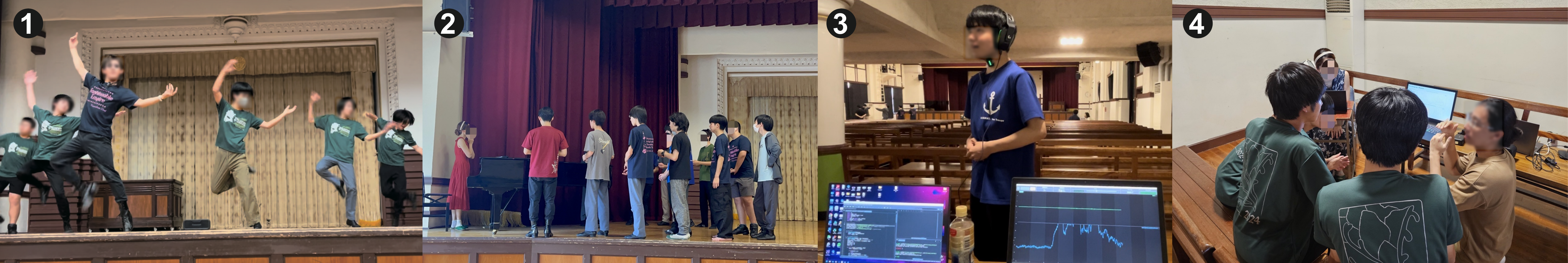}
    \caption{Photos of participants during the experienced singers study: (1) Stage rehearsal for the Phantom musical. (2) Warm-up sessions ensuring participants' proficiency in the selected solfeggio before recording. (3) Participant wearing the sensing setup for the solfeggio recording.(4) The focus group discussion were conducted together with experts, amateurs and researchers.}
    \Description{The figure shows four pictures taken during the experienced singers study. The pictures contain participants during their stage rehearsal, the warm-up, while wearing the sensing setup and during the focus group discussion. All participants' faces were anonymized by blurring.}
    \label{fig:fieldstudy}
\end{figure*}

\begin{table*}[]
    \centering
    \small
    \begin{tabular}{c|c|c|c|c|c|c|c}
        Focus Group & Participants & Expertise & Year & {PerAb $\uparrow$}& {SinAb $\uparrow$} & {RMS Mean (SD) $\uparrow$} & {SPR Mean (SD)$\uparrow$}\\ \hline
         F1 & P1.1 & Tenor & 2 & 41 & 25 & -57.30 (165.18) & -0.248 (0.367)\\
          & P1.2 & Tenor & 1 & 39 & 23 & -75.33 (154.78) & -0.247 (0.367) \\
          & P1.3 & Tenor & 2 & 43 & 34 & -33.77 (111.40) & -0.247 (0.367) \\
          & P1.4 & Tenor & > 10 & 43 & 37 & -33.74 (131.66)  &  -0.246 (0.366)\\ \hline
         F2 & P2.1 & Bass & 2 & 28 & 32 & -125.87 (157.67)  & -0.245 (0.366)\\
          & P2.2 & Bass & 4-5 & 38 & 12 & -27.00 (162.21)  & -0.247 (0.367)\\
          & P2.3 & Bass & 4-5 & 44 & 33 & -  &  -0.246 (0.366)\\
          & P2.4 & Bass & < 1 & 28 & 24 & -88.35 (150.10)  &  -0.244 (0.365)\\ \hline
         F3 & P3.1 & Tenor & 4-5 & 40 & 30 & -95.24 (153.60)  & -0.247 (0.369)\\
          & P3.2 & Bass & 1 & 44 & 30 & -90.01 (159.65) & -0.247 (0.367)\\
          & P3.3 & Tenor & > 10 & 39 & 26 & -76.93 (146.11) & -0.246 (0.367)\\
          & P3.4 & Bass & 4-5 & 40 & 32 & -90.12 (153.37) & -0.244 (0.365)\\ \hline
         F4 & P4.1 & Tenor & 3 & 46 & 24 & -58.34 (175.46)  &  -0.248 (0.367)\\
          & P4.2 & Bass & 1 & 38 & 26 & -58.63 (129.48)  & -0.238 (0.358)\\
          & P4.3 & Bass & 2 & 51 & 33 &-93.59 (155.78)  &  -0.244 (0.366)\\
    \end{tabular}
    \caption{Participant to Focus Group Allocation of our experienced singers study. Demographics are collected by the Goldsmiths Musical Sophistication Index (Gold-MSI)~\cite{mullensiefen2014musicality}: Year (Experience), PerAb (Perceptual Ability), SinAb (Singing Ability). Descriptive Statistics of EMG and SPR are calculated separately for students and expet by the difference of normalized EMG Root Mean Square (RMS) and SPR.}
    \Description{The table gives an overview of the experienced singers study participants' allocation to the focus groups, their expertise, years of experience, perceptual ability, singing ability, as well as mean and standard deviation of EMG RMS and SPR.}
    \label{tab:fg_allo}
\end{table*}

\subsection{Study Procedure}
\label{sec: 6.1 Data Collection}
The study was approved by the local IRB, following a systematic procedure to ensure consistency across data collection:\newline

\noindent\textit{1) Welcome \& Warm-Up:} The study started with a 10 minutes warm-up, guided by a vocal expert with over 10 years of professional musical stage experience. \newline

\noindent\textit{2) MVC Calibration:}
Before the recording trials, participants performed a maximum voluntary contraction (MVC) task by sustaining a vowel (/a:/) at their highest comfortable intensity for about three seconds. This task was repeated multiple times within a 35-second calibration window, with a five-second rest period between each sustained vocalization. The EMG signal was monitored in real time using the Avanti Sensor interface (see the third picture in Figure~\ref{fig:fieldstudy}), which automatically computed the maximum amplitude and baseline noise. This MVC value was used to normalize subsequent EMG data. \newline

\noindent\textit{3) Recording Trials:}
Each participant practiced and mimicked \textit{Christine's audition segment} from the musical \textit{"You Are Music"}~\footnote{The musical notation in Fig~\ref{fig:initial analysis} is an original illustration created by the author based on "You Are Music" by Maury Yeston, © 1991, purchased from Musicnotes.com. This usage falls under the Fair Use doctrine for academic purposes.}. Headphones were used to ensure accurate feedback and prevent interference from ambient noise.\newline

\noindent\textit{4) Feedback Sessions:}
Next, participants were divided into focus groups, where they first received traditional feedback from the vocal expert. Second, they were presented with the EMG and audio analysis of their singing, and finally discussed with their peers about questions stated by the group moderating researcher.

\subsection{Participants \& Recruitment} We recruited 15 male experienced singers with the average age of 15 years (\textit{Min}=13; \textit{Max}=17; \textit{SD}=1.146)), all of whom had received continuous formal vocal training experience for more than one year. Consent of the parents was given before the study. The participants were either bass or tenor singers.

For the focus group discussion, the participants were randomly split into four groups, leading to one tenor-only, one bass-only and two mixed groups. For an overview of the allocation of participants to the focus groups see Table~\ref{tab:fg_allo}.

\subsection{Results}
\label{sec: 6.2 Results}

Before analyzing our data, we had to process it to extract key features, such as EMG Normalized Root Mean Square (EMG RMS) and Singing Power Ratio (SPR). The EMG RMS reflects vocal cord muscles activation~\cite{manabe2003unvoiced,de1997test}, indicating muscle coordination and strength, while SPR~\cite{usha2017objective} measures energy distribution in the voice, enabling accurate tracking of vocal power and resonance across different frequencies.
The Raw EMG data (sampled at 4370Hz) was normalized using the Maximum Voluntary Contraction (MVC)   method~\cite{arjunan2014computation} to allow for consistent muscle activity comparisons. Data were segmented into 200ms windows to calculate Root Mean Square (RMS) values.
For the SPR, microphone array data (sampled at 48kHz) were first processed with a band-pass filter (500--4000 Hz) to suppress low-frequency
rumble and high-frequency noise. Singing Power Ratio (SPR) was then calculated using the Librosa package~\cite{mcfee2015librosa},following standard acoustic
definitions of SPR~\cite{usha2017objective}. Librosa was used for signal loading,
Short-Time Fourier Transform (STFT) computation, and spectral energy extraction. Specifically, we applied \texttt{librosa.stft()} to obtain the magnitude spectrogram, from which SPR was derived as the ratio of energy in the high-frequency band (2--4 kHz) to that in the low-frequency band (0.5--1 kHz). The band energies were computed by summing the squared magnitude values across their respective frequency bins over time.as follows:

\begin{equation}
\begin{split}
    \text{SPR} = 10 \cdot \log_{10}\left(\frac{\int_{2 \text{ kHz}}^{4 \text{ kHz}} P(f) \, df}{\int_{0.5 \text{ kHz}}^{1 \text{ kHz}} P(f) \, df}\right)
\end{split}
\label{eq:spr}
\end{equation}

\subsubsection{\textbf{Sensing:} Tenor Group Shows Higher Muscle Engagement; Expert Displays Strong EMG-SPR Correlation.}
\label{sec:6.2.1 sensing}
\paragraph{Tenor versus Bass}

\begin{figure*}
    \centering
    \includegraphics[width=\textwidth]{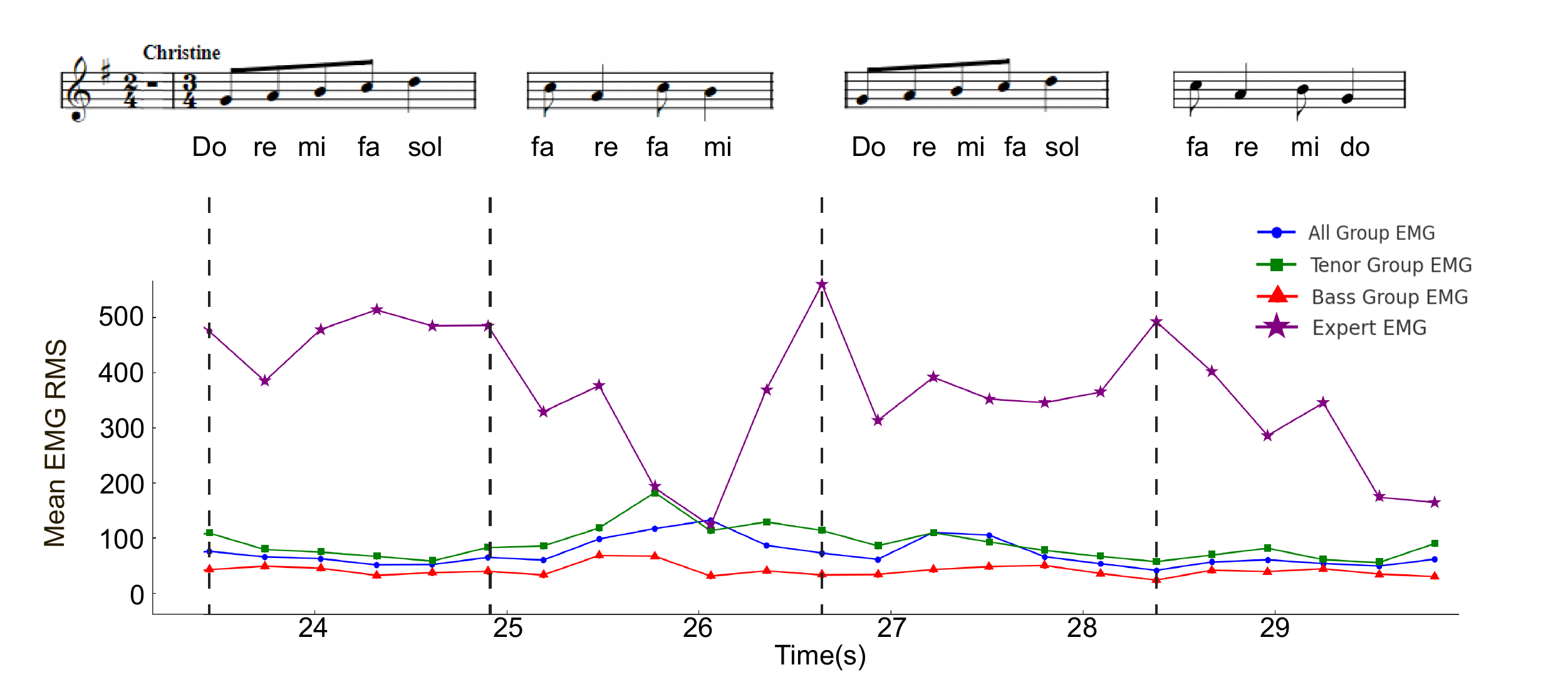}
    \caption{Upper: The song 'You Are Music' has a tempo of 102 BPM, with 3 beats per measure and a beat duration of 0.588 seconds, equivalent to a quarter note or two eighth notes (0.292 seconds). Bottom: The line chart compares EMG RMS values for different participant groups and an expert over the time period from 23s to 30s. It displays the average EMG RMS values for the All Group (blue), Tenor Group (green), Bass Group (red), and the Expert (purple), illustrating differences in muscle activity and performance between the expert and the student groups.}
    \Description{The upper part of the figure shows the musical sheet of 'You are Music'. The lower part of the figure shows a line chart of average EMG RMS values differentiated  by all groups, tenor group, bass group, and expert group.}
    \label{fig:initial analysis}
\end{figure*}

The Tenor Group had RMS Mean differences (RMS and SPR differences were calculated by the student's \textit{Mean Value} minus the expert's) ranging from -33.73 to -95.23, with moderate variability (see Table~\ref{tab:fg_allo}). In contrast, the Bass Group exhibited larger differences (-125.87 to -26.9972) and greater variability, indicating less consistent muscle engagement. Both groups demonstrated stable SPR differences around -0.245, with minimal variability.

The Tenor Group displayed higher and more consistent EMG RMS values, indicating greater muscle engagement, while the Bass Group exhibited wider variability and lower muscle activation. These differences are further illustrated in the group comparison Figure~\ref{fig:initial analysis}, where the tenor group appears to have more controlled muscle activation, closer to expert performance.

Expert EMG demonstrated significantly higher and more flexible RMS values than the trained group, suggesting more controlled and efficient muscle activation. These results align with the dataset findings, reinforcing the expert's superior performance.

\begin{figure*}
    \centering
    \includegraphics[width= 0.9\textwidth]{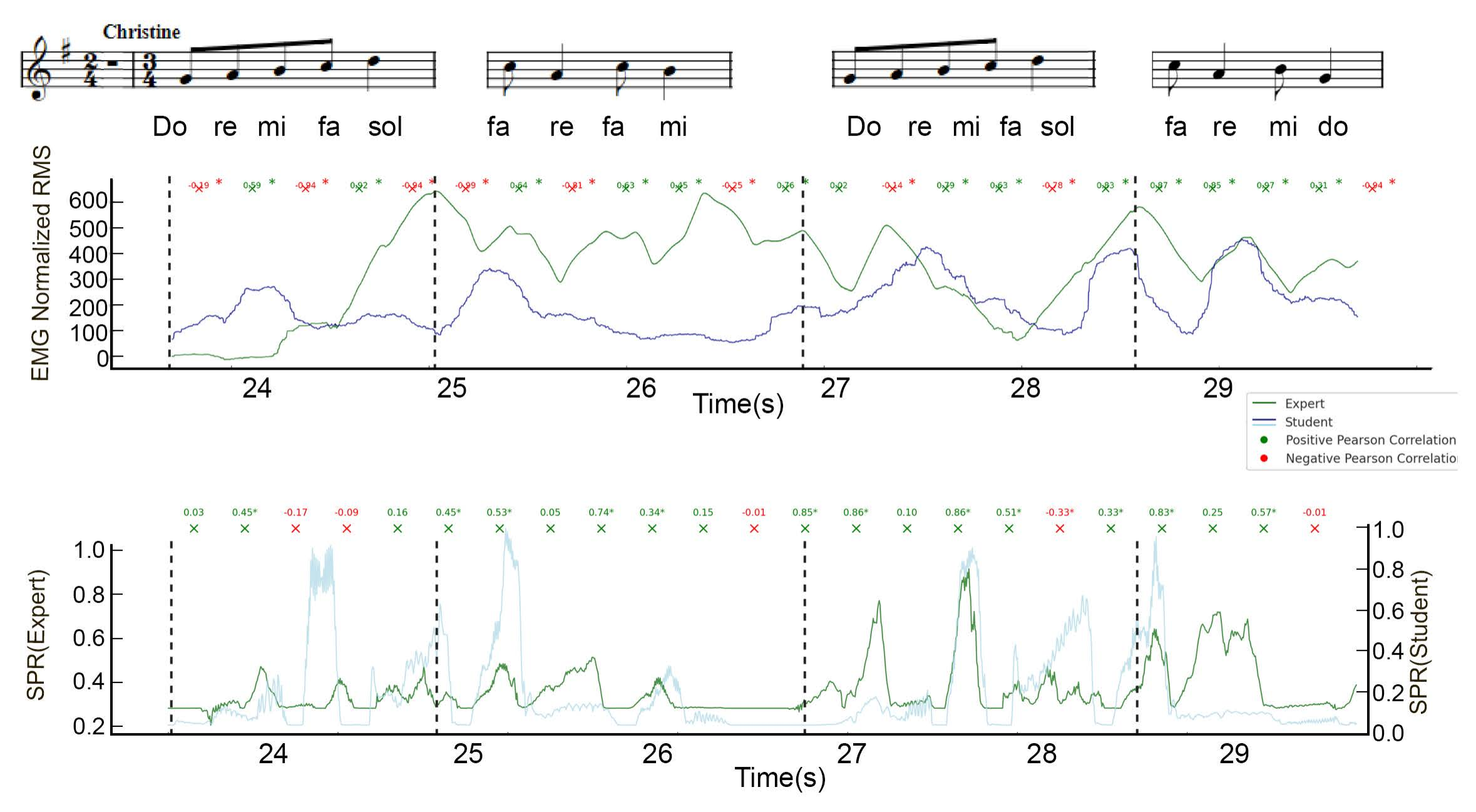}
    \caption{In the period (21s-28s) of the singing segment (upper graph), the trends in EMG and SPR of an expert, and a student who leads the Christine role, are shown. Significance markers (asterisks) are placed above certain correlation values to denote statistical significance.}
    \Description{The figure shows, in the period (21s-28s) of the singing segment，the trends in EMG and SPR of an expert, and a student who leads the Christine role. From the observation of Pearson values, the expert and student show similar acoustic patterns, with more positive correlations, suggesting that their acoustic characteristics align despite differences in physiological responses, indicating similar techniques.}
    \label{fig:initial analysi}
\end{figure*}

\paragraph{EMG versus SPR}
We downsampled EMG RMS and SPR, aligned them with the notes and calculated Pearson correlations across the expert and student participants.
Pearson correlation coefficients between EMG and SPR revealed a strong positive correlation for the expert ($r = .75$, $p < .01$), indicating a significant relationship between vocal muscle activation and singing power.

Specifically, from the observation of Pearson values in Figure~\ref{fig:initial analysi}, we see that the expert and experienced student show similar acoustic patterns, with more positive correlations, suggesting that their acoustic characteristics align despite differences in physiological responses, indicating similar techniques.

Professional participants' correlations varied from approximately $r = .09$ to $r = .53$, showing different levels of engagement with the sensing feedback. This finding highlights a closer relationship between SPR and EMG signals in experts, potentially reflecting higher coordination between vocal performance and muscle control.

\paragraph{Principal Component Analysis (PCA)}
We further applied PCA and statistical tests using Python to explore differences between professional groups (Tenor and Bass) based on three Factors of PerAb (Perceptual Ability), SinAb (Singing Abilities), and RMS Mean. One participant (P2.3) was excluded as the outlier since EMG sensor could not be securely attached due to their moisture skin.

PCA reduced the dataset's dimensionality by extracting two principal components that explained 86.29 \texttt{percent} of the total variance. Principal Component 1 (PC1), which explained 44.26 \texttt{percent} of the variance, was influenced by PerAb  and RMS Mean Difference. Principal Component 2 (PC2), explaining 42.03 \texttt{percent}, was driven by SinAb and RMS Mean Difference. These components effectively capture the variation in perceptual and singing abilities and muscle signals.

In the PCA space, the Tenor group appeared more clustered, indicating consistency among participants, while the Bass group was more spread out, particularly along PC1, suggesting greater individual variability. To test for group differences, we conducted $t$-tests and a Wilcoxon signed-rank test. The $t$-tests showed no significant differences in PerAb ($t(8.64)=0.91$,
$p=.388$, Cohen's $d=0.44$, 95\% CI $[-4.05, 9.44]$), SinAb ($t(12.74)=0.21$, $p=.838$, $d=0.11$, 95\% CI $[-6.37, 7.72]$), or EMG Mean Difference ($t(11.02)=1.40$, $p=.189$,
$d=0.75$, 95\% CI $[-11.68, 52.51]$). The Wilcoxon signed-rank test also did not reveal significant differences.

\subsubsection{\textbf{Focus Group:} Participants appreciated the precision of tech- \\ nology-assisted training but recommended combining it with traditional coaching for clearer feedback.}
\label{sec: 6.2.2 Focus Group}
To analyze the interview data from the focus groups, we first transcribed the interviews into written form. Next, we did an inductive coding. Therefore, first one researcher familiarized themselves with the data and proposed a codebook. Second, two coder applied the codebook to half of the transcripts. Finally, both coder validated each others' assessments, and disagreements were resolved in a review meeting. We followed suggestions by literature, \textit{not} to do multiple independent codings and calculate Inter Coder Reliability (ICR) to prove reliability~\cite{braun2013successful,ortloff2023IRR,mcdonald2019IRR}, acknowledging the influence of the researcher on the process. After this, both researchers grouped the codes into our main themes.

\paragraph{\textbf{Theme 1: Impressions of Technology-Assisted Training}}
When we first explained the visualization to the participants, some struggled to understand it immediately. After answering participants' questions, all participants were confident to gain a even better understanding the more frequent they engage with the visualization. Aspects of the visualization participants were most interested in were (1) the \textsc{time interval}, (2) their applied \textsc{muscle force} represented in the visualization, and (3) \textsc{voice differences} between the ``user'', i.e. male participants, and the female expert-baseline, their singing was compared against.

Participants from F1 and F2 recognized that the visualization showed timing differences between when they started to sing compared to the expert-baseline. P1.1 suggested to use a 0.1 seconds \textsc{time interval} (instead of 0.2 seconds) for visualizing the data even more accurately.

Further, participants showed an interest in how their applied \textsc{muscle force} is represented to see how they have to adjust their own muscle, as P1.3 asked ``\textit{Well, this vertical axis is the amount of muscle power applied, right?}'' (P1.3).
At the same time, the visualization of the \textsc{muscle force} made some participants realize that they could improve their singing by increasing the muscle control.

Finally, participants wondered if the gender of the expert that sang in the training-baseline should match the gender of the person training with the baseline. P1.4 and P1.2 concluded that ``\textit{There are differences between male and female voices.}'' (P1.4), so ``\textit{It's better to do it with the same gender.}'' (P1.2).
Participants from the bass group F2 considered it as helpful for their singing to see the different amplitudes in the graph of female and male singers:
``\textit{The second graph shows the difference between the male and female voices, so the amplitude of their voices is completely different.}'' (P2.1).

\paragraph{\textbf{Theme 2: Limitations of Traditional Vocal Training}}
The participants' discussion about their first impression of the visualization transitioned into discussing their singing voice types (bass or tenor), and how traditional training is limited to address specific aspects of it.

Participants from the tenor group especially experienced challenges with reaching high pitches. They discussed that they often reached a threshold which they cannot overcome,
and that additional difficulties can occur for persons singing in \textit{falsetto}\footnote{Falsetto~\cite{enwiki:1240557845} is the vocal register occupying the frequency range just above the modal voice register and overlapping with it by approximately one octave. It is produced by the vibration of the ligamentous edges of the vocal cords, in whole or in part.}.
Because of this, some participants mentioned to sing ``\textit{based on instinct}'' (P1.4) to achieve best performance, or trying to imitate the original singer of a song.
Participants referred to this as ``intuitive training'', meaning that they do not follow a dedicated training schedules or routine, but rather sing intuitively during their day, as P3.1 said: ``\textit{I sing in the bath}'' (P3.1).

In contrast, other participants mentioned to have dedicated training routines including listening to others singing a song before learning it themselves,
singing karaoke,
doing certain vocal exercises,
or recording themselves singing to listen to it.
Additionally, P3.4 mentioned to sing during his workouts to include various body muscles in his singing: ``\textit{I was told that if I push a desk while singing, I can learn how to use my stomach, so I'm doing it.}'' (P3.4).
However, solely relying on auditory feedback was perceived as less effective to improve the learners' vocal \textit{Performance} compared to training with a technology-assisted method as also found in our novice study (see Section \ref{sec:5 NOVICE USER STUDY}).

Within their traditional training, participants experienced challenges in controlling their natural and singing voice,
finding a balance between their normal voice and falsetto,
and memorizing their training without visual cues that they can re-check at a later time.

\paragraph{\textbf{Theme 3: Advantages of Technology-Assisted Training}}
Our participants identified several advantages of training with our visualization compared to traditional training.
First, they considered the visualization to be more precise, as it directly points out which parts of the singing need improvement.
This makes it easier for participants to ``see'' the volume and the pitch of their voice at one glance.
Further, participants reported that this helps them to easier understand on which parts of their muscle they need to focus: ``\textit{So, when I use such a graph, it is easier to understand the specifics from my muscles.}''(P1.4).
Participants also felt that the visualization was beneficial when it came to language barriers between them and their traditional coaches.
Further, participants liked that they were able to confirm their expectation of singing with the visualization. Participants from the bass group compared the amplitude of their singing with the tenor group: The bass groups' visualization had less fluctuation compared to the tenor group, which is
in line with expectations when comparing bass and tenor singers. Participants therefore expressed a feeling of relief when identifying this behavior with the visualization: ``\textit{I'm in the bass group, so I'm high stability that's how it is. That's good. I'm relieved.}'' (P2.1). Similar observations were made in the mixed group F4. There, participants discussed the difference between tenor and bass voices, comparing the different types to the expert-baseline.

\paragraph{\textbf{Theme 4: Concerns towards Technology-Assisted Training}}
Our participants not only considered advantages of technology-assisted training but also expressed some concerns.
Participants considered that singing coaches need to set some expert singing to be the standard-baseline for singing when they employ the visualization feedback in their current training. This led to the concern that students might not be able to fulfill these criteria anymore. Further, they wondered whether the perception of ``good'' singing might change, once there is a technical solution to evaluate someone's singing: ``\textit{When the machine thinks it's good, everyone thinks it's good.}'' (P4.3), only leaving little room for subjective interpretation.
Participants further missed the human interaction with their coach, and
P1.4 argued that interacting only with a numerical representation in the visualization is not as natural, since ``\textit{A human being is not a machine. So, it is not something that can be moved by inputting numerical values. In the end, we use our nerves, our senses, and our bodies, so I think it's a little difficult to understand even if you look at it numerically.}'' (P1.4).

\paragraph{\textbf{Theme 5: Future of Technology-Assisted Training}}
Our participants came up with ideas how they would like to incorporate the technology-assisted training in their current routines. The participants' ideas to \textbf{include the visualization as it is} are (1) to get an overview of their progress, (2) as a substitute for finding a learning group, (3) as a substitute for a coach when training at home, or (4) as an addition when training karaoke.

However, participants also had several suggestions on how to combine the visualization with traditional training which we detail below.
When comparing traditional feedback from their coach to the feedback provided by the visualization, participants considered the form of feedback: While their coach can provide them feedback combined with suggestions how to improve their singing, with the visualization participants had to find out by trying themselves what they need to do in order to improve.
Hence, participants expressed a need to \textbf{supplement the visualization by advice}, similar to the advice a coach would give.

In contrast to supplementing the visualization by advice, P1.4 suggested to \textbf{use the visualization as a supplement to traditional feedback}.
Participants argued that it is hard to understand the correlation between the visualization and muscle movement solely from the visualization. Making participants suggest that a connection between what they see and feel is necessary, and could be achieved by discussing the visualization with their coach during traditional training.

Overall, the participants expressed a preference to \textbf{use a mixture of traditional feedback and the new visualization} method, as P1.2 stated ``\textit{I don't think it is right to choose only one of them.}'' (P1.2).

\subsubsection{Result Summary: RQ3}
Traditional feedback currently encounters several challenges in terms of availability when no coach is present, language-barriers between coach and learner, and precision, e.g., when learners have to review recordings of their singing, leading to delayed inaccurate feedback. Despite some concerns how technology-assisted training could lead to higher expectations and singing standards, all our focus groups found positive aspects and opportunities on how to employ it. Participants therefore came up with ideas on how to combine traditional and technology-assisted training in their current routines.
\vspace{5mm}

\section{DISCUSSION}
\label{sec: 7 DISCUSSION}
Our three studies demonstrate that EMG and ultrasonography provide complementary feedback modalities for vocal training, with effectiveness varying by skill level and voice type. Study 1 established that these sensing technologies can differentiate muscle activation patterns between novice and experienced singers. Study 2 showed that novices improved vocal performance with both technologies, though each imposed distinct cognitive demands. Study 3 revealed that experienced singers exhibited strong coordination between muscle activation and acoustic output, with notable differences between tenor and bass voice types. Below, we synthesize these findings to address how physiological sensing can augment vocal pedagogy.

\subsection{Summary of Key Study Results}
\label{sec: 7.1 Effectiveness of Vocal Sensing}
Both EMG and UI demonstrated promising capabilities in detecting and differentiating vocal muscle activity across different skill levels.

In our second user study with novice singers, EMG enhanced the participants' perceived control, even though the stability decreased post-training, while UI improved vocal cord length control but required more effort to interpret. The decrease of the stability likely is a temporary effect that can have two possible explanations. First, this temporary decline should not be interpreted as performance loss, but rather as a transitional adaptation phase. Novices are required to recruit and coordinate muscles in a way that differs from intuitive singing, and such re-coordination is often accompanied
by short-term instability. Prior vocal pedagogy literature similarly notes that increased or misdirected tension can momentarily degrade sound quality when new techniques are introduced~\cite{miller1996structure}. Our findings align with this
view: the sensing feedback drew attention to specific muscle groups, prompting novices to consciously engage muscles they previously controlled implicitly,
resulting in short-term variability.
Second, novices need to ``re-learn'' their ability to control their vocal cords in a new way compared to singing by intuition. More long-term training lets individuals regain this stability. As shown in our first study, novices typically have less stability in their EMG signals compared to more experienced singers.

In our third study, (the experienced singers study) we revealed a strong correlation between EMG and singing power in experts, with Tenor participants showing more consistent muscle engagement than Bass participants. While participants appreciated the precision of the sensing feedback, they also expressed concerns about over-reliance on numerical data and emphasized the importance of integrating technology with traditional vocal training for a more balanced approach.

Our studies collectively showed that both EMG and UI provide valuable insights into vocal muscle control, but their impact on perception differs. EMG offers intuitive, real-time feedback, improving muscle engagement and control, while UI provides detailed visual cues for vocal cord movement, helpful in early-stage training or refining specific techniques. The need for such visual feedback was also evident in our final study with the experienced singers where we investigated EMG exclusively.
Our first study revealed significant differences between beginners and experts, with EMG excelling in temporal stability and UI in visualizing muscle activity. The novice study highlighted a trade-off between perceived control and cognitive overload with EMG, while UI required more effort to interpret but improved vocal control. The experienced singers study further demonstrated that tenor singers exhibited better muscle engagement, and a strong correlation between EMG and SPR was observed in expert singers, indicating higher coordination between muscle activity and vocal performance.

\subsection{Enhancing Muscle Control and Vocal Performance}
\label{sec: 7.3 Skill Transfer-ability}
\subsubsection{Traditional versus Technical}
\label{sec: 7.3.1 Traditional versus Technical}
One current challenge of singing training revealed by our experienced singers study is that frequent training is necessary, but often no learning groups for extracurricular learning (without coach) are available. This leads to singers currently rely solely on auditory feedback from recordings of their own voice.
But results of our novice study showed that learners can reach a greater improvement of their vocal \textit{Performance} when utilizing UI or EMG as training methods.
\subsubsection{Tenor versus Bass}
\label{sec: 7.3.2 Tenor versus Bass}
In comparing EMG and UI for skill transferability, our findings showed that EMG stability decreased post-training for novices, especially in lower pitches, while UI improved vocal cord length control, particularly in higher pitches. In the professional group, tenors exhibited more consistent muscle control than bass singers, who showed greater variability and difficulty in skill transfer. Pearson correlations between EMG and SPR highlighted stronger coordination in the expert, with students, especially bass singers, showing lower correlations. PCA analysis confirmed that tenors had more consistent performance, while bass singers displayed higher variability, indicating the need for more targeted feedback to improve muscle stability. Overall, EMG benefits tenors more, while UI aids novices in muscle coordination, with bass singers requiring additional support.

\subsection{Implicating the Design Guideline for Musical Training and Performance}
\label{sec: 7.4 Design Considerations for Training System}
In designing effective vocal training systems, the integration of EMG and UI offers significant advantages by providing detailed feedback on muscle activity and vocal cord movements. EMG excels in capturing fine motor control in vocal cords, offering precise data for vocal performance, while UI provides direct visual cues, helping users understand complex vocal mechanisms.

Building on advances in Human-Computer Interaction for music performance, such as BrainiBeats~\cite{10.1145/3544549.3585910}, these technologies can enhance not only vocal pedagogy but also applications in speech therapy and personalized music education. However, usability challenges with UI, especially as a wearable, highlight the need for refining its design for broader accessibility. Combining these sensing technologies into user-friendly, real-time systems could support continuous feedback, improving training for users at all skill levels.

\subsection{Limitations and Future Works}
\label{limitations}
\paragraph{\textbf{Longer-term Training Periods}}
One limitation we identified is that evaluating pitch immediately after training may not fully reflect long-term effects. To explore this, we recalled 6 novice participants to assess their performance on "Little Star." The EMG method showed average improvements of 8.16 Hz in pitch accuracy and 7.53 Hz in consistency~\cite{mcfee2015librosa}, while the UI method had smaller gains. These results suggest both methods enhance pitch control, with EMG showing stronger effects. Future work should focus on how sensory feedback can support long-term improvements in muscle control and vocal performance.

\paragraph{\textbf{Interface Design}}
We recognized the cognitive load in interpreting raw EMG and UI data in the novice study and the need for improved visual feedback in the experienced singers study. Future systems should offer more intuitive feedback aligned with expert models and extended training routines. Additionally, current sensing is limited by Bluetooth interference between the EMG sensor and ultrasound probe, which we aim to resolve with a dual-modal interface and automated UI using neural networks.

\paragraph{\textbf{Group Variability and Sample Size Limitations}}
Although PCA analysis revealed different distributions between the Tenor and Bass groups in the principal component space, the statistical tests failed to show significant group differences. This may be attributed to the small sample size or high within-group variability, indicating that further research with a larger sample size is needed to explore these potential differences in more depth.

\paragraph{\textbf{Gender and Developmental Confounds}}
In Study 3, our male adolescent participants (ages 13--17) were compared against a female expert baseline. We chose this baseline because the musical piece (``You Are Music'') is written for female soprano, but this creates a confound: male and female singers have different laryngeal anatomy, so differences in EMG and SPR (Figure~\ref{fig:initial analysis}) may reflect sex rather than skill level. Our adolescent participants were also undergoing vocal development, adding further variability. Future work should use sex- and age-matched baselines to better isolate training effects.

\section{CONCLUSION}
In conclusion, this research explored the vocal cord muscles activities using novel sensing (EMG and UI) for vocal performance assessment. Through our user studies with novice, experienced and expert singers (44 in total), we demonstrated that these sensing technologies can effectively capture vocal muscle activity and provide valuable feedback for improving vocal control. While both methods showed promise in enhancing pitch accuracy and stability, future work is needed to optimize user experience and investigate long-term training effects. Our findings highlight the potential of integrating advanced sensing technologies into vocal training, offering more precise, intuitive feedback to support diverse user needs in vocal education and performance.

\begin{acks}
This work is conducted under the IoT Accessibility Toolkit Project supported by JST Presto Grant Number JPMJPR2132 and funded by JST SPRING H09GQ24152, collaborating with Koike lab at the Institute of Science Tokyo under the fundings by JST CRONOS JPMJCS24N8.
\end{acks}

\bibliographystyle{ACM-Reference-Format}
\bibliography{CHI_Vocals}

@inproceedings{ortloff2023IRR, 
author = {Ortloff, Anna-Marie and Fassl, Matthias and Ponticello, Alexander and Martius, Florin and Mertens, Anne and Krombholz, Katharina and Smith, Matthew},
title = {Different Researchers, Different Results? Analyzing the Influence of Researcher Experience and Data Type During Qualitative Analysis of an Interview and Survey Study on Security Advice},
year = {2023},
isbn = {9781450394215},
publisher = {Association for Computing Machinery},
address = {New York, NY, USA},
url = {https://doi.org/10.1145/3544548.3580766     } ,
doi = {10.1145/3544548.3580766 },
booktitle = {Proceedings of the 2023 CHI Conference on Human Factors in Computing Systems},
articleno = {864},
numpages = {21},
location = {Hamburg, Germany},
series = {CHI '23}
}

@inproceedings{meng2025placebo,
  title={A Placebo Concert: The Placebo Effect for Visualization of Physiological Audience Data during Experience Recreation in Virtual Reality},
  author={Meng, Xiaru and Ju, Yulan and Kim, Christopher Changmok and He, Yan and Barbareschi, Giulia and Minamizawa, Kouta and Kunze, Kai and Hoppe, Matthias},
  booktitle={Proceedings of the 2025 CHI Conference on Human Factors in Computing Systems},
  pages={1--16},
  year={2025}
}

@article{han2022linking,
  title={Linking Audience Physiology to Choreography},
  author={Han, Jiawen and Chernyshov, George and Sugawa, Moe and Zheng, Dingding and Hynds, Danny and Furukawa, Taichi and Padovani, Marcelo and Minamizawa, Kouta and Marky, Karola and Ward, Jamie A and others},
  journal={ACM Transactions on Computer-Human Interaction},
  year={2022},
  publisher={ACM}
}

@inproceedings{meng2023towards,
  title={Towards enhancing a recorded concert experience in virtual reality by visualizing the physiological data of the audience},
  author={Meng, Xiaru and He, Yan and Kunze, Kai},
  booktitle={Proceedings of the Augmented Humans International Conference 2023},
  pages={330--333},
  year={2023}
}

@article{he2022frisson,
  title={Frisson waves: exploring automatic detection, triggering and sharing of aesthetic chills in music performances},
  author={He, Yan and Chernyshov, George and Han, Jiawen and Zheng, Dingding and Thomsen, Ragnar and Hynds, Danny and Liu, Muyu and Yang, Yuehui and Ju, Yulan and Pai, Yun Suen and others},
  journal={Proceedings of the ACM on Interactive, Mobile, Wearable and Ubiquitous Technologies},
  volume={6},
  number={3},
  pages={1--23},
  year={2022},
  publisher={ACM New York, NY, USA}
}

@inproceedings{chen2024novel,
  title={Novel Sensing Methods for Vocal Technique Analysis: Evaluation on Electromyography and Ultrasonography},
  author={Chen, Kanyu and Wu, Erwin and Saito, Daichi and Peng, Yichen and Liao, Chen-Chieh and Kato, Akira and Koike, Hideki and Kunze, Kai},
  booktitle={2024 IEEE International Symposium on Mixed and Augmented Reality Adjunct (ISMAR-Adjunct)},
  pages={121--125},
  year={2024},
  organization={IEEE}
}

@inproceedings{chen2025exploring,
  title={Exploring Singing Breath: Physiological Insights and Directions for Breath-Aware Augmentation in Mixed Reality Design},
  author={Chen, Kanyu and Chang, Zhuang and Zou, Qianyuan and Kunze, Kai},
  booktitle={Companion of the 2025 ACM International Joint Conference on Pervasive and Ubiquitous Computing},
  pages={702--706},
  year={2025}
}

@inproceedings{chen2025multimodal,
  title={A Multimodal Wearable Sensing System for Vocal Muscle Biofeedback in Singing Pitch Training},
  author={Chen, Kanyu and Kato, Akira and Kunze, Kai},
  booktitle={Companion of the 2025 ACM International Joint Conference on Pervasive and Ubiquitous Computing},
  pages={449--454},
  year={2025}
}

@article{mcdonald2019IRR,
author = {McDonald, Nora and Schoenebeck, Sarita and Forte, Andrea},
title = {Reliability and Inter-rater Reliability in Qualitative Research: Norms and Guidelines for CSCW and HCI Practice},
year = {2019},
issue_date = {November 2019},
publisher = {Association for Computing Machinery},
address = {New York, NY, USA},
volume = {3},
number = {CSCW},
url = {https://doi.org/10.1145/3359174     } , 
doi = {10.1145/3359174 },
journal = {Proc. ACM Hum.-Comput. Interact.},
month = {nov},
articleno = {72},
numpages = {23}
}

@book{braun2013successful,
  title={{Successful Qualitative Research: A Practical guide for Beginners}},
  author={Braun, Virginia and Clarke, Victoria},
  year={2013},
  publisher={SAGE Publications},
  address = {London}
}

@article{naseth2012constructing,
  title={Constructing the voice: Present and future considerations of vocal pedagogy},
  author={Naseth, Andrew},
  journal={The Choral Journal},
  volume={53},
  number={2},
  pages={39},
  year={2012},
  publisher={American Choral Directors Association}
}

@article{teixeira2015acoustic,
  title={Acoustic analysis of vocal dysphonia},
  author={Teixeira, Jo{\~a}o Paulo and Fernandes, Paula Odete},
  journal={Procedia Computer Science},
  volume={64},
  pages={466--473},
  year={2015},
  publisher={Elsevier}
}

@article{Steinhauer2019,
   author = {Kimberly M Steinhauer and Mary McDonald Klimek},
   doi = {10.1080/23268263.2019.1605707},
   issue = {3},
   journal = {Voice and Speech Review},
   pages = {354-359},
   publisher = {Routledge},
   title = {Vocal Traditions: Estill Voice Training®},
   volume = {13},
   url = {https://doi.org/10.1080/23268263.2019.1605707},
   year = {2019},
}

@article{mcclellan2011comparative,
  title={A comparative analysis of speech level singing and traditional vocal training in the United States},
  author={McClellan, Josef William},
  year={2011}
}

@article{Sundberg2017,
   author = {Johan Sundberg and Maddalena Bitelli and Annika Holmberg and Ville Laaksonen},
   doi = {https://doi.org/10.1016/j.jvoice.2017.02.009},
   issn = {0892-1997},
   issue = {5},
   journal = {Journal of Voice},
   pages = {528-535},
   title = {The “Overdrive” Mode in the “Complete Vocal Technique”: A Preliminary Study},
   volume = {31},
   url = {https://www.sciencedirect.com/science/article/pii/S0892199716305094},
   year = {2017},
}

@article{titze2002rules,
  title={Rules for controlling low-dimensional vocal fold models with muscle activation},
  author={Titze, Ingo R and Story, Brad H},
  journal={The Journal of the Acoustical Society of America},
  volume={112},
  number={3},
  pages={1064--1076},
  year={2002},
  publisher={Acoustical Society of America}
}

@article{belyk2018does,
  title={How does human motor cortex regulate vocal pitch in singers?},
  author={Belyk, Michel and Lee, Yune S and Brown, Steven},
  journal={Royal Society Open Science},
  volume={5},
  number={8},
  pages={172208},
  year={2018},
  publisher={The Royal Society}
}

@article{buchthal1959electromyography,
  title={Electromyography of intrinsic laryngeal muscles},
  author={Buchthal, Fritz},
  journal={Quarterly Journal of Experimental Physiology and Cognate Medical Sciences: Translation and Integration},
  volume={44},
  number={2},
  pages={137--148},
  year={1959},
  publisher={Wiley Online Library}
}

@article{gay1972electromyography,
  title={Electromyography of the intrinsic laryngeal muscles during phonation},
  author={Gay, Thomas and Strome, Marshall and Hirose, Hajime and Sawashima, Masayuki},
  journal={Annals of Otology, Rhinology \& Laryngology},
  volume={81},
  number={3},
  pages={401--409},
  year={1972},
  publisher={SAGE Publications Sage CA: Los Angeles, CA}
}

@article{kempster1988effects,
  title={Effects of electrical stimulation of cricothyroid and thyroarytenoid muscles on voice fundamental frequency},
  author={Kempster, Gail B and Larson, Charles R and Kistler, Michael K},
  journal={Journal of Voice},
  volume={2},
  number={3},
  pages={221--229},
  year={1988},
  publisher={Elsevier}
}

@inproceedings{kimura2019sottovoce,
  title={SottoVoce: An ultrasound imaging-based silent speech interaction using deep neural networks},
  author={Kimura, Naoki and Kono, Michinari and Rekimoto, Jun},
  booktitle={Proceedings of the 2019 CHI Conference on Human Factors in Computing Systems},
  pages={1--11},
  year={2019}
}

@article{zhu2022towards,
  title={Towards evaluating pitch-related phonation function in speech communication using high-density surface electromyography},
  author={Zhu, Mingxing and Wang, Xin and Deng, Hanjie and He, Yuchao and Zhang, Haoshi and Liu, Zhenzhen and Chen, Shixiong and Wang, Mingjiang and Li, Guanglin},
  journal={Frontiers in Neuroscience},
  volume={16},
  pages={941594},
  year={2022},
  publisher={Frontiers}
}

@article{vojtech2021surface,
  title={Surface electromyography--based recognition, synthesis, and perception of prosodic subvocal speech},
  author={Vojtech, Jennifer M and Chan, Michael D and Shiwani, Bhawna and Roy, Serge H and Heaton, James T and Meltzner, Geoffrey S and Contessa, Paola and De Luca, Gianluca and Patel, Rupal and Kline, Joshua C},
  journal={Journal of Speech, Language, and Hearing Research},
  volume={64},
  number={6S},
  pages={2134--2153},
  year={2021},
  publisher={ASHA}
}

@inproceedings{reed2020surface,
  title={Surface electromyography for direct vocal control},
  author={Reed, Courtney and McPherson, Andrew and others},
  year={2020},
  organization={International Conference on New Interfaces for Musical Expression (NIME)}
}

@inproceedings{reed2022singing,
  title={Singing knit: soft knit biosensing for augmenting vocal performances},
  author={Reed, Courtney N and Skach, Sophie and Strohmeier, Paul and McPherson, Andrew P},
  booktitle={Proceedings of the Augmented Humans International Conference 2022},
  pages={170--183},
  year={2022}
}

@inproceedings{Cotton2021Body,
	author = {Cotton, Kelsey and Sanches, Pedro and Tsaknaki, Vasiliki and Karpashevich, Pavel},
	booktitle = {NIME 2021},
	year = {2021},
	month = {apr 29},
	note = {https://nime.pubpub.org/pub/ntm5kbux},
	organization = {},
	title = {The {Body} {Electric}: A {NIME} designed through and with the somatic experience of singing},
}

@book{cohen1995time,
  title={Time-frequency analysis},
  author={Cohen, Leon},
  volume={778},
  year={1995},
  publisher={Prentice hall New Jersey}
}

@book{oppenheim1999discrete,
  title={Discrete-time signal processing},
  author={Oppenheim, Alan V},
  year={1999},
  publisher={Pearson Education India}
}

@inproceedings{farrus2007jitter,
  title={Jitter and shimmer measurements for speaker recognition},
  author={Farr{\'u}s, Mireia and Hernando, Javier and Ejarque, Pascual},
  booktitle={8th Annual Conference of the International Speech Communication Association; 2007 Aug. 27-31; Antwerp (Belgium).[place unknown]: ISCA; 2007. p. 778-81.},
  year={2007},
  organization={International Speech Communication Association (ISCA)}
}

@article{kumar2017vocal,
  title={Vocal cord dysfunction: Ultrasonography-aided diagnosis during routine airway examination},
  author={Kumar, Amarjeet and Sinha, Chandni and Singh, Akhilesh Kumar and Bhadani, Umesh Kumar},
  journal={Saudi Journal of Anaesthesia},
  volume={11},
  number={3},
  pages={370--371},
  year={2017},
  publisher={Medknow}
}

@article{soas2017,
author = {Tapal, Adam and Oren, Ela and Dar, Reuven and Eitam, Baruch},
year = {2017},
month = {09},
pages = {1552},
title = {The Sense of Agency Scale: A Measure of Consciously Perceived Control over One's Mind, Body, and the Immediate Environment},
volume = {8},
journal = {Frontiers in Psychology},
doi = {10.3389/fpsyg.2017.01552}
}

@article{hart1988development,
  author = {Hart, Sandra G and Staveland, Lowell E},
  journal = {Human mental workload},
  number = 3,
  pages = {139--183},
  publisher = {Amsterdam, Holland},
  title = {Development of NASA-TLX (Task Load Index): Results of empirical and theoretical research},
  volume = 1,
  year = 1988
}

@ARTICLE{Sarto2021-zb,
  title     = "Implementing ultrasound imaging for the assessment of muscle and
               tendon properties in elite sports: Practical aspects,
               methodological considerations and future directions",
  author    = "Sarto, Fabio and Sp{\"o}rri, J{\"o}rg and Fitze, Daniel P and
               Quinlan, Jonathan I and Narici, Marco V and Franchi, Martino V",
  journal   = "Sports Med.",
  publisher = "Springer Science and Business Media LLC",
  volume    =  51,
  number    =  6,
  pages     = "1151--1170",
  month     =  jun,
  year      =  2021,
  copyright = "https://creativecommons.org/licenses/by/4.0",
  language  = "en"
}

@ARTICLE{Nagae2023-ql,
  title     = "Muscle ultrasound and its application to point-of-care
               ultrasonography: a narrative review",
  author    = "Nagae, Masaaki and Umegaki, Hiroyuki and Yoshiko, Akito and
               Fujita, Kosuke",
  journal   = "Ann. Med.",
  publisher = "Informa UK Limited",
  volume    =  55,
  number    =  1,
  pages     = "190--197",
  month     =  dec,
  year      =  2023,
  copyright = "http://creativecommons.org/licenses/by/4.0/",
  language  = "en"
}

@ARTICLE{Paris2021-ck,
  title     = "Muscle composition analysis of ultrasound images: A narrative
               review of texture analysis",
  author    = "Paris, Michael T and Mourtzakis, Marina",
  journal   = "Ultrasound Med. Biol.",
  publisher = "Elsevier BV",
  volume    =  47,
  number    =  4,
  pages     = "880--895",
  month     =  apr,
  year      =  2021,
  language  = "en"
}

@inproceedings{lopes2015,
author = {Lopes, Pedro and Ion, Alexandra and Baudisch, Patrick},
title = {Impacto: Simulating Physical Impact by Combining Tactile Stimulation with Electrical Muscle Stimulation},
year = {2015},
isbn = {9781450337793},
publisher = {Association for Computing Machinery},
address = {New York, NY, USA},
url = {https://doi.org/10.1145/2807442.2807443},
doi = {10.1145/2807442.2807443},
booktitle = {Proceedings of the 28th Annual ACM Symposium on User Interface Software \& Technology},
pages = {11–19},
numpages = {9},
location = {Charlotte, NC, USA},
series = {UIST '15}
}

@inproceedings{takahashi20221,
author = {Takahashi, Akifumi and Brooks, Jas and Kajimoto, Hiroyuki and Lopes, Pedro},
title = {Increasing Electrical Muscle Stimulation's Dexterity by means of Back of the Hand Actuation},
year = {2021},
isbn = {9781450380966},
publisher = {Association for Computing Machinery},
address = {New York, NY, USA},
url = {https://doi.org/10.1145/3411764.3445761},
doi = {10.1145/3411764.3445761},
booktitle = {Proceedings of the 2021 CHI Conference on Human Factors in Computing Systems},
articleno = {216},
numpages = {12},
location = {Yokohama, Japan},
series = {CHI '21}
}

@inproceedings{rudolph2022,
author = {Rudolph, Julius Cosmo Romeo and Holman, David and De Araujo, Bruno and Jota, Ricardo and Wigdor, Daniel and Savage, Valkyrie},
title = {Sensing Hand Interactions with Everyday Objects by Profiling Wrist Topography},
year = {2022},
isbn = {9781450391474},
publisher = {Association for Computing Machinery},
address = {New York, NY, USA},
url = {https://doi.org/10.1145/3490149.3501320},
doi = {10.1145/3490149.3501320},
booktitle = {Proceedings of the Sixteenth International Conference on Tangible, Embedded, and Embodied Interaction},
articleno = {14},
numpages = {14},
location = {Daejeon, Republic of Korea},
series = {TEI '22}
}

@ARTICLE{Esposito2018-mu,
  title    = "A piezoresistive sensor to measure muscle contraction and
              mechanomyography",
  author   = "Esposito, Daniele and Andreozzi, Emilio and Fratini, Antonio and
              Gargiulo, Gaetano D and Savino, Sergio and Niola, Vincenzo and
              Bifulco, Paolo",
  journal  = "Sensors (Basel)",
  volume   =  18,
  number   =  8,
  month    =  aug,
  year     =  2018,
  language = "en"
}

@inproceedings{Zhang2016,
author = {Zhang, Yang and Xiao, Robert and Harrison, Chris},
title = {Advancing Hand Gesture Recognition with High Resolution Electrical Impedance Tomography},
year = {2016},
isbn = {9781450341899},
publisher = {Association for Computing Machinery},
address = {New York, NY, USA},
url = {https://doi.org/10.1145/2984511.2984574},
doi = {10.1145/2984511.2984574},
booktitle = {Proceedings of the 29th Annual Symposium on User Interface Software and Technology},
pages = {843–850},
numpages = {8},
location = {Tokyo, Japan},
series = {UIST '16}
}

@ARTICLE{Zheng2021,
  author={Zheng, Enhao and Wan, Jiacheng and Yang, Lin and Wang, Qining and Qiao, Hong},
  journal={IEEE Robotics and Automation Letters}, 
  title={Wrist Angle Estimation With a Musculoskeletal Model Driven by Electrical Impedance Tomography Signals}, 
  year={2021},
  volume={6},
  number={2},
  pages={2186-2193},
  doi={10.1109/LRA.2021.3060400}}

@inproceedings{Zhu2021,
author = {Zhu, Junyi and Snowden, Jackson C and Verdejo, Joshua and Chen, Emily and Zhang, Paul and Ghaednia, Hamid and Schwab, Joseph H and Mueller, Stefanie},
title = {EIT-kit: An Electrical Impedance Tomography Toolkit for Health and Motion Sensing},
year = {2021},
isbn = {9781450386357},
publisher = {Association for Computing Machinery},
address = {New York, NY, USA},
url = {https://doi.org/10.1145/3472749.3474758},
doi = {10.1145/3472749.3474758},
booktitle = {The 34th Annual ACM Symposium on User Interface Software and Technology},
pages = {400–413},
numpages = {14},
location = {Virtual Event, USA},
series = {UIST '21}
}

@inproceedings{Zhu2022,
author = {Zhu, Junyi and Lei, Yuxuan and Shah, Aashini and Schein, Gila and Ghaednia, Hamid and Schwab, Joseph and Harteveld, Casper and Mueller, Stefanie},
title = {MuscleRehab: Improving Unsupervised Physical Rehabilitation by Monitoring and Visualizing Muscle Engagement},
year = {2022},
isbn = {9781450393201},
publisher = {Association for Computing Machinery},
address = {New York, NY, USA},
url = {https://doi.org/10.1145/3526113.3545705},
doi = {10.1145/3526113.3545705},
booktitle = {Proceedings of the 35th Annual ACM Symposium on User Interface Software and Technology},
articleno = {33},
numpages = {14},
location = {Bend, OR, USA},
series = {UIST '22}
}

@inproceedings{Eddy2023,
author = {Eddy, Ethan and Scheme, Erik J and Bateman, Scott},
title = {A Framework and Call to Action for the Future Development of EMG-Based Input in HCI},
year = {2023},
isbn = {9781450394215},
publisher = {Association for Computing Machinery},
address = {New York, NY, USA},
url = {https://doi.org/10.1145/3544548.3580962},
doi = {10.1145/3544548.3580962},
booktitle = {Proceedings of the 2023 CHI Conference on Human Factors in Computing Systems},
articleno = {145},
numpages = {23},
location = {Hamburg, Germany},
series = {CHI '23}
}

@inproceedings{Saisho2019,
author = {Saisho, Osamu and Tsukada, Shingo and Nakashima, Hiroshi and Imamura, Hiroshi and Takaori, Kyoichi},
title = {Enhancing support for optimal muscle usage in sports: coaching and skill-improvement tracking with sEMG},
year = {2019},
isbn = {9781450368704},
publisher = {Association for Computing Machinery},
address = {New York, NY, USA},
url = {https://doi.org/10.1145/3341163.3347722},
doi = {10.1145/3341163.3347722},
booktitle = {Proceedings of the 2019 ACM International Symposium on Wearable Computers},
pages = {206–210},
numpages = {5},
location = {London, United Kingdom},
series = {ISWC '19}
}

@inproceedings{Papakostas2019,
author = {Papakostas, Michalis and Kanal, Varun and Abujelala, Maher and Tsiakas, Konstantinos and Makedon, Fillia},
title = {Physical fatigue detection through EMG wearables and subjective user reports: a machine learning approach towards adaptive rehabilitation},
year = {2019},
isbn = {9781450362320},
publisher = {Association for Computing Machinery},
address = {New York, NY, USA},
url = {https://doi.org/10.1145/3316782.3322772},
doi = {10.1145/3316782.3322772},
booktitle = {Proceedings of the 12th ACM International Conference on PErvasive Technologies Related to Assistive Environments},
pages = {475–481},
numpages = {7},
location = {Rhodes, Greece},
series = {PETRA '19}
}

@article{usha2017objective,
  title={Objective identification of prepubertal female singers and non-singers by singing power ratio using matlab},
  author={Usha, M and Geetha, YV and Darshan, YS},
  journal={Journal of Voice},
  volume={31},
  number={2},
  pages={157--160},
  year={2017},
  publisher={Elsevier}
}

@article{desjardins2022respiratory,
  title={Respiratory muscle strength training to improve vocal function in patients with presbyphonia},
  author={Desjardins, Maude and Halstead, Lucinda and Simpson, Annie and Flume, Patrick and Bonilha, Heather Shaw},
  journal={Journal of Voice},
  volume={36},
  number={3},
  pages={344--360},
  year={2022},
  publisher={Elsevier}
}

@book{welch2019oxford,
  title={The Oxford handbook of singing},
  author={Welch, Graham F and Howard, David M and Nix, John},
  year={2019},
  publisher={Oxford University Press}
}

@article{de1997test,
  title={Test-retest study of the GRBAS scale: influence of experience and professional background on perceptual rating of voice quality},
  author={De Bodt, Marc S and Wuyts, Floris L and Van de Heyning, Paul H and Croux, Christophe},
  journal={Journal of voice},
  volume={11},
  number={1},
  pages={74--80},
  year={1997},
  publisher={Elsevier}
}

@article{luck2018comparison,
  title={A comparison of written, vocal, and video feedback when training teachers},
  author={Luck, Kally M and Lerman, Dorothea C and Wu, Wai-Ling and Dupuis, Danielle L and Hussein, Louisa A},
  journal={Journal of Behavioral Education},
  volume={27},
  pages={124--144},
  year={2018},
  publisher={Springer}
}

@article{song2013assessment,
  title={Assessment of vocal cord function and voice disorders},
  author={Song, Phillip},
  journal={Principles and practice of interventional pulmonology},
  pages={137--149},
  year={2013},
  publisher={Springer}
}

@inproceedings{manabe2003unvoiced,
  title={Unvoiced speech recognition using EMG-mime speech recognition},
  author={Manabe, Hiroyuki and Hiraiwa, Akira and Sugimura, Toshiaki},
  booktitle={CHI'03 extended abstracts on Human factors in computing systems},
  pages={794--795},
  year={2003}
}

@article{miller1996structure,
  title={The structure of singing: System and art in vocal technique},
  author={Miller, Richard},
  journal={(No Title)},
  year={1996}
}

@article{lee2022real,
  title={Real-time feedback system for classical singing training using singing power ratio},
  author={Lee, Jihun and Kim, Hyun and Park, Soyeon},
  journal={Journal of Voice},
  volume={36},
  number={1},
  pages={129--136},
  year={2022},
  publisher={Elsevier}
}

@article{sundberg2020singing,
  title={Singing power ratio and vocal health in professional classical singers},
  author={Sundberg, Johan},
  journal={Journal of Voice},
  volume={34},
  number={5},
  pages={690--696},
  year={2020},
  publisher={Elsevier}
}

@article{allen2019advanced,
  title={Advanced waveform analysis for improving the quality of vocal recordings},
  author={Allen, John B},
  journal={Journal of the Acoustical Society of America},
  volume={146},
  number={4},
  pages={2457--2468},
  year={2019},
  publisher={Acoustical Society of America}
}

@article{huang2021enhanced,
  title={Enhanced waveform analysis for detecting vibrato and pitch modulation in vocal performances},
  author={Huang, Peng and Li, Yuan and Wang, Zheng},
  journal={Journal of the Acoustical Society of America},
  volume={150},
  number={3},
  pages={1753--1762},
  year={2021},
  publisher={Acoustical Society of America}
}

@inproceedings{10.1145/3544549.3583824,
author = {Lee, Jaejun and Cho, Hyeyoon and Kim, Yonghyun},
title = {Bean Academy: A Music Composition Game for Beginners with Vocal Query Transcription},
year = {2023},
isbn = {9781450394222},
publisher = {Association for Computing Machinery},
address = {New York, NY, USA},
url = {https://doi.org/10.1145/3544549.3583824},
doi = {10.1145/3544549.3583824},
booktitle = {Extended Abstracts of the 2023 CHI Conference on Human Factors in Computing Systems},
articleno = {591},
numpages = {6},
location = {Hamburg, Germany},
series = {CHI EA '23}
}

@inproceedings{10.1145/2468356.2468435,
author = {Krause, Markus and Smeddinck, Jan and Meyer, Ronald},
title = {A digital game to support voice treatment for parkinson's disease},
year = {2013},
isbn = {9781450319522},
publisher = {Association for Computing Machinery},
address = {New York, NY, USA},
url = {https://doi.org/10.1145/2468356.2468435},
doi = {10.1145/2468356.2468435},
booktitle = {CHI '13 Extended Abstracts on Human Factors in Computing Systems},
pages = {445–450},
numpages = {6},
location = {Paris, France},
series = {CHI EA '13}
}

@inproceedings{10.1145/2468356.2479485,
author = {Shi, Yang and Yang, Cheng},
title = {Celestia: a vocal interaction music game},
year = {2013},
isbn = {9781450319522},
publisher = {Association for Computing Machinery},
address = {New York, NY, USA},
url = {https://doi.org/10.1145/2468356.2479485},
doi = {10.1145/2468356.2479485},
booktitle = {CHI '13 Extended Abstracts on Human Factors in Computing Systems},
pages = {2647–2650},
numpages = {4},
location = {Paris, France},
series = {CHI EA '13}
}

@inproceedings{10.1145/3544549.3585910,
author = {Ceccato, Caterina and Pruss, Ethel and Vrins, Anita and Prinsen, Jos and Alimardani, Maryam},
title = {BrainiBeats: A dual brain-computer interface for musical composition using inter-brain synchrony and emotional valence},
year = {2023},
isbn = {9781450394222},
publisher = {Association for Computing Machinery},
address = {New York, NY, USA},
url = {https://doi.org/10.1145/3544549.3585910},
doi = {10.1145/3544549.3585910},
booktitle = {Extended Abstracts of the 2023 CHI Conference on Human Factors in Computing Systems},
articleno = {56},
numpages = {7},
location = {Hamburg, Germany},
series = {CHI EA '23}
}

@inproceedings{10.1145/2468356.2479620,
author = {Morreale, Fabio and Masu, Raul and De Angeli, Antonella and Rota, Paolo},
title = {The music room},
year = {2013},
isbn = {9781450319522},
publisher = {Association for Computing Machinery},
address = {New York, NY, USA},
url = {https://doi.org/10.1145/2468356.2479620},
doi = {10.1145/2468356.2479620},
booktitle = {CHI '13 Extended Abstracts on Human Factors in Computing Systems},
pages = {3099–3102},
numpages = {4},
location = {Paris, France},
series = {CHI EA '13}
}

@inproceedings{10.1145/2468356.2479596,
author = {Unander-Scharin, Carl and H\"{o}\"{o}k, Kristina and Elblaus, Ludvig},
title = {The throat III: disforming operatic voices through a novel interactive instrument},
year = {2013},
isbn = {9781450319522},
publisher = {Association for Computing Machinery},
address = {New York, NY, USA},
url = {https://doi.org/10.1145/2468356.2479596},
doi = {10.1145/2468356.2479596},
booktitle = {CHI '13 Extended Abstracts on Human Factors in Computing Systems},
pages = {3007–3010},
numpages = {4},
location = {Paris, France},
series = {CHI EA '13}
}

@inproceedings{10.1145/2556288.2557050,
author = {Unander-Scharin, Carl and Unander-Scharin, \r{A}sa and H\"{o}\"{o}k, Kristina},
title = {The vocal chorder: empowering opera singers with a large interactive instrument},
year = {2014},
isbn = {9781450324731},
publisher = {Association for Computing Machinery},
address = {New York, NY, USA},
url = {https://doi.org/10.1145/2556288.2557050},
doi = {10.1145/2556288.2557050},
booktitle = {Proceedings of the SIGCHI Conference on Human Factors in Computing Systems},
pages = {1001–1010},
numpages = {10},
location = {Toronto, Ontario, Canada},
series = {CHI '14}
}

@inproceedings{10.1145/2559206.2574798,
author = {Unander-Scharin, Carl and Unander-Scharin, \r{A}sa and H\"{o}\"{o}k, Kristina and Elblaus, Ludvig},
title = {Interacting with the vocal chorder: re-empowering the opera diva},
year = {2014},
isbn = {9781450324748},
publisher = {Association for Computing Machinery},
address = {New York, NY, USA},
url = {https://doi.org/10.1145/2559206.2574798},
doi = {10.1145/2559206.2574798},
booktitle = {CHI '14 Extended Abstracts on Human Factors in Computing Systems},
pages = {603–606},
numpages = {4},
location = {Toronto, Ontario, Canada},
series = {CHI EA '14}
}

@inproceedings{10.1145/3544548.3581162,
author = {Schlagowski, Ruben and Nazarenko, Dariia and Can, Yekta and Gupta, Kunal and Mertes, Silvan and Billinghurst, Mark and Andr\'{e}, Elisabeth},
title = {Wish You Were Here: Mental and Physiological Effects of Remote Music Collaboration in Mixed Reality},
year = {2023},
isbn = {9781450394215},
publisher = {Association for Computing Machinery},
address = {New York, NY, USA},
url = {https://doi.org/10.1145/3544548.3581162},
doi = {10.1145/3544548.3581162},
booktitle = {Proceedings of the 2023 CHI Conference on Human Factors in Computing Systems},
articleno = {102},
numpages = {16},
location = {Hamburg, Germany},
series = {CHI '23}
}

@article{yamauchi2010perceptual,
  title={Perceptual evaluation of pathological voice quality: A comparative analysis between the RASATI and GRBASI scales},
  author={Yamauchi, Emi Juliana and Imaizumi, Satoshi and Maruyama, Hagino and Haji, Tomoyuki},
  journal={Logopedics Phoniatrics Vocology},
  volume={35},
  number={3},
  pages={121--128},
  year={2010},
  publisher={Taylor \& Francis}
}

@misc{ enwiki:1240557845,
    author = "{Wikipedia contributors}",
    title = "Falsetto --- {Wikipedia}{,} The Free Encyclopedia",
    year = "2024",
    url = "https://en.wikipedia.org/w/index.php?title=Falsetto&oldid=1240557845",
    note = "[Online; accessed 12-September-2024]"
  }

@inproceedings{mcfee2015librosa,
  title={librosa: Audio and music signal analysis in python.},
  author={McFee, Brian and Raffel, Colin and Liang, Dawen and Ellis, Daniel PW and McVicar, Matt and Battenberg, Eric and Nieto, Oriol},
  booktitle={SciPy},
  pages={18--24},
  year={2015}
}

@article{arjunan2014computation,
  title={Computation and evaluation of features of surface electromyogram to identify the force of muscle contraction and muscle fatigue},
  author={Arjunan, Sridhar P and Kumar, Dinesh K and Naik, Ganesh},
  journal={BioMed research international},
  volume={2014},
  number={1},
  pages={197960},
  year={2014},
  publisher={Wiley Online Library}
}

@article{mullensiefen2014musicality,
  title={The musicality of non-musicians: An index for assessing musical sophistication in the general population},
  author={M{\"u}llensiefen, Daniel and Gingras, Bruno and Musil, Jason and Stewart, Lauren},
  journal={PloS one},
  volume={9},
  number={2},
  pages={e89642},
  year={2014},
  publisher={Public Library of Science San Francisco, USA}
}

@book{cohen2013statistical,
  title={Statistical power analysis for the behavioral sciences},
  author={Cohen, Jacob},
  year={2013},
  publisher={routledge}
}

@article{lakens2013calculating,
  title={Calculating and reporting effect sizes to facilitate cumulative science: a practical primer for t-tests and ANOVAs},
  author={Lakens, Dani{\"e}l},
  journal={Frontiers in psychology},
  volume={4},
  pages={863},
  year={2013},
  publisher={Frontiers Media SA}
}

@article{benjamini1995controlling,
  title={Controlling the false discovery rate: a practical and powerful approach to multiple testing},
  author={Benjamini, Yoav and Hochberg, Yosef},
  journal={Journal of the Royal statistical society: series B (Methodological)},
  volume={57},
  number={1},
  pages={289--300},
  year={1995},
  publisher={Wiley Online Library}
}

@book{gelman2007data,
  title={Data analysis using regression and multilevel/hierarchical models},
  author={Gelman, Andrew and Hill, Jennifer},
  year={2007},
  publisher={Cambridge university press}
}

\appendix

\end{document}